\documentclass[aps,prb,twocolumn]{revtex4-1}
\usepackage{amssymb}
\usepackage{amsmath,bm}
\usepackage{graphicx,color}
 \usepackage{epstopdf}
\usepackage{epsfig}
\usepackage{undertilde}
\usepackage{multirow}
\usepackage[cal=pxtx]{mathalfa}
\usepackage{accents}
\DeclareMathAccent{\wtilde}{\mathord}{largesymbols}{"65}

\usepackage{stackengine}
\stackMath
\newcommand\tenq[2][1]{%
	\def\useanchorwidth{T}%
	\ifnum#1>1%
	\stackunder[0pt]{\tenq[\numexpr#1-1\relax]{#2}}{\scriptscriptstyle\sim}%
	\else%
	\stackunder[1pt]{#2}{\scriptscriptstyle\sim}%
	\fi%
}

\newcommand{\B}[1]{{\underline{#1}}} 

\begin{document}

\title{Micro-mechanical insights into the stress transmission in strongly aggregating colloidal gel }
\author{Divas Singh Dagur$^1$, Yezaz Ahmed Gadi Man$^1$, Saikat Roy$^{1,}$}
\email{Corresponding author: saikat.roy@iitrpr.ac.in}
\affiliation{$^1$ Department of Chemical Engineering, IIT Ropar, Rupnagar, Punjab, India 140001 }

\begin{abstract}
 Predicting the mechanical response of the soft gel materials under external deformation is of paramount importance in many areas, such as foods, pharmaceuticals, solid-liquid separations, cosmetics, aerogels and drug delivery. Most of the understanding of the elasticity of gel materials is based on the concept of fractal scaling with very little microscopic insights. Previous experimental observations strongly suggest that the gel material loses the fractal correlations upon deformation and the range of packing fraction up to which the fractal scaling can be applied is very limited. Also, correctly implementing the fractal modeling requires identifying the elastic backbone, which is a formidable task. So far, there is no clear understanding of the gel elasticity at high packing fraction and the correct length scale that governs its mechanical response. In this work, we undertake extensive numerical simulations to elucidate the different aspects of stress transmission in the gel materials. We observe the existence of two percolating networks of compressive and tensile normal forces close to the gel point. We also find that the probability distribution for the compressive and tensile part normalized by their respective mean shows universal behavior irrespective of varied interaction potential, thermal energy and particle size distribution. Interestingly, there are also large number of contacts with zero normal force, and consequently, a peak in the normal force distribution is observed at $f_n\approx0$ even at higher pressure. We also identify the critical internal state parameters such as mean normal force, force anisotropies and the average coordination number and propose simple constitutive relations that relate the different components of the stress to the internal state parameters. The agreement between our model prediction and the simulation observation is excellent. It is shown that the anisotropy in the force networks gives rise to the normal stress difference in the soft gel materials. Our results strongly demonstrate that the mechanical response of the gel system is governed mainly by the particle length scale phenomena with a complex interplay between compressive and tensile forces at the particle contact.

\end{abstract}

\maketitle
\section{Introduction}
Colloidal suspensions can form amorphous solids in many different ways \cite{zaccarelli2007colloidal} depending on the inter-particle potential, thermal energy, volume fraction, and external stress. Colloidal glasses are formed at high volume fractions due to the dynamical arrest arising from crowded interactions and caging effects. In contrast, gels form a system spanning rigid percolating network at very low particle packing fractions and manifest solid-like properties. The fundamental physics behind the origin of elasticity in diverse amorphous soft solids is far from understood compared to crystalline solids where elasticity emerges from broken symmetry. One of the intriguing questions is: `` How does such a low packing fraction disordered gel manifest solid-like properties, and how can its solidity be related to other amorphous solids formed at high packing fractions?'' At the gel point, a space spanning percolating network (connectivity percolation) is not sufficient to guarantee solid-like properties; one needs the rigidity percolation additionally to ensure elasticity in the network. The rigidity of the gel network demands the structural organization of the network in such a way that the mechanical balance is established sufficiently both at the particle and macroscopic levels  \cite{dagur2022spatial}. Previous experiments and simulations \cite{hsiao2012role,tsurusawa2019direct} showed that the mechanical stability of the colloidal gel stems from  the isotropic percolation of isostatic rigid clusters. All these studies were performed on the colloidal gel formed via depletion interaction with centrosymmetric forces whereas in reality nanoscale surface roughness can lead to frictional contacts and substantial tangential interactions as shown explicitly in previous experiments \cite{pantina2005elasticity,furst2007yielding}. Frictional interactions will lead to the breakdown of the isostaticity concept in understanding the emergence of elasticity in the gel network. As per Maxwell's constraint counting arguments, the average coordination number, $z$ has to be $2d$ ($d$ is the Euclidean dimension) for a network to be isostatic and rigid for central pairwise interaction. In the presence of rolling and sliding resistance, the gel network's rigidity can be achieved at a very low average coordination number due to the force indeterminacy generated as a consequence of the preparation protocol.

Due to its unique mechanical and transport properties, gel finds extensive applications in the industry such as food, pharmaceutical, cosmetics, and everyday life. From the application point of view, predicting the mechanical response of the colloidal gel under external deformation is of extreme importance. In this regard, the colloidal gel network is considered to be composed of somewhat ambiguous fractal blobs \cite{kantor1984elastic,shih1990scaling} of some characteristic length scale upto which the system shows fractal behaviour. The fractal flocs are assumed to be the principal load bearing blocks of the percolating network and control the elastic response of the system. Considering stretching and bending contributions to the elastic energy, the elastic modulus of the network can be related to the average floc size and upon using the fractal scaling of the floc size with the volume fraction, the evolution of the modulus and other relevant parameters with the volume fraction can be predicted. Recent works \cite{roy2016b,roy2020micro,roy2016universality} extended the fractal model and developed a constitutive relation to predict the stresses in the consolidating gel network under uniaxial deformation. Although the prediction of the axial stress was in line with the experimental observations     \cite{buscall1988scaling}, recent measurements \cite{islam2021normal} point to the failure of the theory to predict the transverse stress and, consequently the normal stress difference in the gel system. The failure of the theory can be attributed to its blindness to account for the internal stresses that are present in the gel network, even without any confining stress. The basic premise upon which the theory is erected, i.e interaction force is zero at each contact and the strain is measured from a stress free state, is incorrect as we will show. Gel network can form a self-sustained equilibrium structure by a delicate balance between compressive and tensile forces and the residual stresses can not be neglected in the limit of large particle modulus. The absence of a standard zero-load reference state precludes the formulation of any constitutive relations between stress and strain in a traditional way. Moreover, almost all existing models use the fractal scaling even at high packing fraction to predict the gel rheology, which is questionable as shown in the previous works \cite{seto2013compressive}. So there is no definitive picture of the gel elasticity at high packing fraction and the correct length scale that controls the mechanical response.

In this study, we carry out detailed numerical simulations to understand the nature of stress transmission in the soft gel materials which support tensile and compressive forces as well as finite rolling torque. It is observed that the propagation of stress through the gel materials is non-trivial and happens via two co-existing percolating networks:compressive and tensile close to the gel point. At higher pressure, the compressive network grows stronger whereas the tensile forces become localized and are no more percolating. The probability distributions of the tensile and compressive forces normalized by their respective mean shows universal features regardless of different interaction potential, thermal energy, size distribution. One of the notable facets of the stress transmission in gel materials is the presence of large number of contacts with the zero normal force unlike the jammed granular materials where most of the normal forces lie close to the mean. We also identify the crucial state parameters such as mean normal force, force anisotropies, average coordination number, which play a pivotal role in the gel rheology. We propose a simple model that relates the different stress components to the internal state parameters and find excellent agreement with the simulation data. It is also shown that the anisotropy in the force networks gives rise to the normal stress difference in soft gel materials. Surprisingly, our results strongly indicate that the mechanical response of the gel network is predominantly dictated by the particle length scale phenomena with a non-trivial interplay between the compressive and tensile forces at the particle contact.  This observation is in direct contradiction with the existing literature where a fractal characteristic length scale is considered to describe the elasticity in gel materials even though the network is not fractal at higher packing fractions. Thus our results present a novel picture of the stress transmission in the gel systems, which can describe the mechanical response both at low and high packing fractions in a unified manner.
\section{Simulation of colloidal gel}

We perform simulation of uniaxial compression of flocculated colloidal gel system using open source codes, Large-scale Atomic/Molecular Massively Parallel Simulator (LAMMPS) \cite{plimpton1995fast}.The spatial and temporal evolution of the colloidal gel network will be dictated mainly by the elastic and hydrodynamic forces. In this study, the strain rates ($\approx 10^{-5} s^{-1}$) are kept very small to mimic quasistatic compression and to eliminate inertial effects. In the limit of slow compression, the ratio of the hydrodynamic force ($\eta \dot{\epsilon}$) to the elastic force on a particle becomes negligible ($G\dot{\epsilon}$). Here, $\eta$, $\dot{\epsilon}$, $G$ and $\epsilon$ denote solvent viscosity, strain rate, particle modulus and strain respectively. Prior experimental observations  \cite{brambilla2011highly,lin2015hydrodynamic} also point towards the dominance of the network stress over the viscous stress in the slow strain rate regime once a system spanning network has been formed. In view of these experimental observations and simple order of magnitude analysis of the relevant forces, we do not consider hydrodynamic forces in the simulation. 

Previous simulations on the colloidal gel systems mostly consider centrosymmetric potential, occasionally with an orientation dependent three-body term \cite{colombo2014stress} to represent the angular rigidity of the interparticle bonds. However, due to the presence of surface roughness, the contact between colloidal particles seems inevitable and consequently the tangential interactions such as sliding and rolling effects will become crucial in the mechanical response as also observed in the experiments \cite{pantina2005elasticity}. To our knowledge, nonhydrodynamic interparticle friction has not been recognized in previous works as a key ingredient in properly representing the colloidal gel interactions at close distances. Our simulation methodology also presents a novel approach to study the gel systems taking into account sliding and rolling resistance that are present due to the nanoscale roughness. It was also recently recognized    \cite{seto2013discontinuous,hsu2018roughness,mari2015discontinuous,lin2015hydrodynamic,singh2020shear} that the contact between particles is essential to reproduce the discontinuous and continuous shear thickening in colloidal suspensions and employed granular (Hookean) interactions on top of the hydrodynamics, Brownian and other relevant interactions. In our simulation also, we used granular interactions at close contact and the Brownian forces which arises due to the interaction with an implicit solvent. Our previous simulations using the same model correctly captured the internal stress field, structure and correlations, which led to the successful reproduction of the macroscopic rheological response in terms of shear and compressive yield stress, normal stress difference in the colloidal gel systems \cite{roy2016b,roy2016universality,roy2016yielding,roy2020micro,islam2021consolidation,islam2021normal,islam2021superposed,seto2013compressive}. In this study, we also model the colloidal gel system by employing discrete element method (DEM) simulation generally applied for granular materials. Note that, our simulation is completely different from the simulation generally employed for the cohesive granular materials that are under no thermal fluctuation.
The simulation employs $N=4000$ soft elastic discs of diameter $D$ placed randomly in a two-dimensional square box of size $150D$ , corresponding to a packing fraction, $\phi\approx0.1$. The contact deformation is modeled as Hookean, the normal part of the elastic repulsive force between a particle pair $i$ and $j$ is given as $\B F_{N}^{e,ij}=-k_{n}\delta \B n_{ij}$ where $k_n$ is the normal stiffness and $\B n_{ij}$ is the center to center unit normal vector. This force acts when the overlap distance between two particles, $\delta < 0$. We also performed simulations with Hertzian interaction where normal elastic force is modeled as $\B F_{N}^{e,ij}=-k_{n}\delta^{3/2}(D/4)^{1/2}\B n_{ij}$. In order to obtain the static equilibrium in reasonable time, normal viscous damping,$-\gamma_n \B {v}_{n_{ij}}$ (here $ \B {v}_{n_{ij}}$ is the normal component of the relative velocity and 
$\gamma_n$ is the normal viscoelastic damping coefficient ) is also added. Similarly for the tangential mode of deformation, the force, $F_{T}^{ij}$ is considered to vary linearly with the tangential overlap till the initiation of sliding which takes place when $F_{T}^{ij}>=\pm\mu F_{N}^{e,ij}$ where $\mu$ is the sliding friction coefficient. Damping is not provided for the tangential part. A constant attractive force of magnitude $F_0$ acts center to center between two particles when a contact is made. Note that the Coulomb inequality is applicable to the repulsive elastic normal force only. The rolling resistance between the particles is taken into account and modeled just like the contact elasticity and friction for the sliding mode. The rolling pseudo-force is given as \cite{luding2008cohesive},
\begin{equation}
\B F_{r}^{ij}=k_r\B \delta_{roll}-\gamma_r \B v_{roll}
\end{equation} 
 where $k_r$ is the rolling stiffness, $\B \delta_{roll}$ is the rolling displacement, $\gamma_r$ is the damping constant for the rolling mode, $\B v_{roll}$ is the  relative rolling velocity. Similar to the sliding friction, the rolling force is limited by $\mu_r F_{N}^{e,ij}$, where $\mu_r$ denotes the rolling friction coefficient. We also run few simulation without any rolling resistance to create different microstructure. In this work, $k_n=k_t=2\times10^5 N/m$, $k_r=k_t/4$, $\mu=\mu_r=0.5$, $\gamma_n=500 kg/s, \gamma_r=50 kg/s$ and $F_0/k_nD=0.0025$. 
The particles are provided random thermal kicks to its translational and rotational degrees of freedom compensated by a damping term. The Langevin equation is solved to update the particle positions and velocities,
\begin{eqnarray}
m_i \frac{d^2 \B x_i}{dt^2} =\B F_i -m_i\gamma_t \frac{d\B x_i}{dt} +\B f_i(t) ;\\
I_i \frac{d^2 \B \theta_i}{dt^2} =\B T_i -I_i\gamma_r \frac{d\B \theta_i}{dt} +\B L_i(t) 
\end{eqnarray}
where $\B f_i(t)$ and $\B L_i(t)$ are respectively a $\delta$-correlated random force/torque having zero mean with the following properties:
\begin{equation}
\langle \B f_i(t) \cdot \B f_j(t+\tau)\rangle = 2 \Gamma \delta (\tau) \delta_{ij};\langle \B L_i(t) \cdot \B L_j(t+\tau)\rangle = 2 \Gamma \delta (\tau) \delta_{ij}
\end{equation}
Here $\gamma_t$ and $\gamma_r $ denote translational and rotational damping coefficient, $m_i$ and $I_i$ are the mass and moment of inertia of the particle $i$, $\B F_i$ and $\B T_i$  are the force and torque due to interparticle interactions, $\Gamma$ is the strength of fluctuations and $\B x_i$ and $\B \theta_i$  denotes the translational and rotational degrees of freedom.Thermal fluctuations will lead to the aggregation of particles as the interaction is attractive and the flocs are brought closer by the uniaxial compression effected by the movement of top and bottom walls simultaneously at a constant speed, $V_0$. Walls are composed of monolayer of particles with the same properties as that of the particles that formed the gel network. The periodic boundary condition is applied on the sides. We employ compression on top of the thermal fluctuations for the fast formation of the percolating network. In fact, as per our simulation protocol, the gel formation process involves two time scales, one is the compression time scale, and another is the Brownian time scale. Fast compression will give rise to higher gelation concentration and dense microstructure, whereas slow compression will lead to open microstructure and lower gel point.

In our simulation, we tune the timescale of the Brownian motion by changing the strength of the thermal fluctuations to obtain different microstructure to see the effect of the preparation history on the properties of interest. We use three different values of $V/V_0$($5.9,18.6$ and $58.7$) where $V$ denotes the characteristic velocity due to the thermal fluctuations and is given as $\sqrt{k_BT/m}$.  Once the flocs connect to form a system spanning percolating network at the gel point, Brownian motion has little effect on the dynamics since the attractive potential is much stronger than the thermal force (Strong gel). The dynamics of the strong gel is determined by the internal stress field induced during the structural arrest at the gel point and the imposed deformation field. When the thermal fluctuations are comparable to the attraction, weak gels are formed and the Brownian forces will play a significant role in the microscopic dynamics of the gel network. We focus only on the strong gels in this study. 
The gel network is compressed uniaxially and quasistatically at very slow strain rate to different target pressures. The system is then allowed to reach mechanical equilibrium i.e. total force and torque on each particle is negligibly small, $O(10^{-7})$ with almost zero kinetic energy, $O(10^{-16})$ at different target pressures before we start measuring the properties of interest. For each pressure, we performed simulations with ten different independent initial configurations.

\section{Stress-force-fabric relationship}
The classical macroscopic Cauchy stress tensor, $\sigma_{ij}$ for any assembly under static equilibrium can be expressed in terms of volume average of the dyadic products between the contact forces, $f_j^c$ and the contact vector, $l_i^c$ (vector connecting center of the particle to the contact point) \cite{landau1986theory,rothenburg1989analytical}:
\begin{equation}
\sigma_{ij}=n_c\langle l_i^c f_j^c \rangle
\end{equation}
where $n_c$ denotes contact density and $i$, $j$ refer to Cartesian coordinates. This expression is valid for any interaction potential and evidently shows that the stress of a static system is intricately related to its internal state described by the set $\{l_i^c, f_j^c\}$. In the limit of large system size, the internal state can be represented statistically by a continuous joint probability density function, $P(l_i,f_j)$ of contact force and contact vector and subsequently the stress tensor becomes:
\begin{equation}
\sigma_{ij}=n_c\iint l_if_jP(l_i,f_j) \,dl_i\,df_j
\end{equation}
For  monodisperse circular discs, $\B l= (D/2)\B n$, where $\B n$ is the unit contact normal and contact force in the local reference frame ($\B n, \B t$) is given as, $f_j=f_n n_j+f_t t_j$, where $\B t$ is an unit vector orthogonal to $\B n$, $f_n$ and $f_t$ are respectively the normal and tangential components of the contact force. Upon neglecting the correlations between contact force and contact vector, the stress tensor for a two dimensional system is given as \cite{rothenburg1989analytical},
\begin{equation}
\sigma_{ij}=\frac{n_cD}{2}\int_{0}^{2\pi} \left(\bar{f_n}(\theta)n_in_j+\bar{f_t}(\theta)n_it_j\right) P(\theta) \,d\theta \ \label{stress}
\end{equation}
where $P(\theta)$ is the probability density function of the orientations $\theta$ (measured with respect to $x$ axis) of the contact normals and $\bar{f_n}(\theta)$, $\bar{f_t}(\theta)$ represent average normal and tangential forces for a group of contacts with the orientation $\theta$. Furthermore, due to $\pi$ periodicity, $P(\theta)$, $\bar{f_n}(\theta)$, and $\bar{f_t}(\theta)$ can be approximated by the following second order Fourier expansions :

\begin{align}
 P(\theta)&\approx \frac{1}{2\pi}\left\lbrace 1+a\, cos\,2 \left(\theta-\theta_a\right) \right\rbrace \label{Ptheta} \\
 \bar{f_n}(\theta)&\approx \bar{f}_m \left\lbrace 1+a_n\, cos\,2\left(\theta-\theta_f\right) \right\rbrace \label{fn} \\
 \bar{f_t}(\theta)&\approx \bar{f}_m a_t\, sin\,2\left(\theta-\theta_t\right) \label{ft}
\end{align}

Here, $a$ represents the anisotropy in contact orientations, $\theta_a$ denotes the corresponding direction of anisotropy, $\bar{f}_m $ is the mean normal force over all contacts, $a_n$ and $a_t$ define the anisotropy in the directional variations of normal and tangential forces with $\theta_f$ and $\theta_t$ being the preferred directions of contact forces. The calculation of these parameters from the discrete simulation data is performed by introducing three second order tensors: $\Phi_{ij}\approx\frac{1}{N_c}\sum n_i n_j$, $\xi_{ij}\approx\frac{1}{N_g}\sum \bar{f_n} n_i n_j$ and $\chi_{ij}\approx\frac{1}{N_g}\sum \bar{f_t} n_i t_j$, where $N_g$ denotes the number of orientation intervals spanning from $0$ to $2\pi$, $N_c$ represents  the total number of contacts. Anisotropy parameters, $a$, $a_n$ and $a_t$ and their preferred orientations are related to the invariants of $\Phi_{ij}$, $\xi_{ij}$, $\chi_{ij}$ and their principal directions:

\begin{align}
&a=\frac{2\sqrt{(\Phi_{11}-\Phi_{22})^2+4\Phi_{12}^2}}{\Phi_{11}+\Phi_{22}} ;\; tan\,2\theta_a=\frac{2\Phi_{12}}{\Phi_{11}-\Phi_{22}}&\\
&a_n=\frac{2\sqrt{(\xi_{11}-\xi_{22})^2+4\xi_{12}^2}}{\xi_{11}+\xi_{22}} ;\; tan\,2\theta_f=\frac{2\xi_{12}}{\xi_{11}-\xi_{22}}\label{an}&\\
&a_t=\frac{2\sqrt{(\chi_{11}-\chi_{22})^2+4\chi_{12}^2}}{\xi_{11}+\xi_{22}} ;\; tan\,2\theta_t=\frac{2\chi_{12}}{\chi_{11}-\chi_{22}}&\\
&\bar{f}_m=\xi_{11}+\xi_{22} &
\end{align}

\begin{figure*}[htbp]
\begin{center}
\centering \includegraphics[scale=0.12]{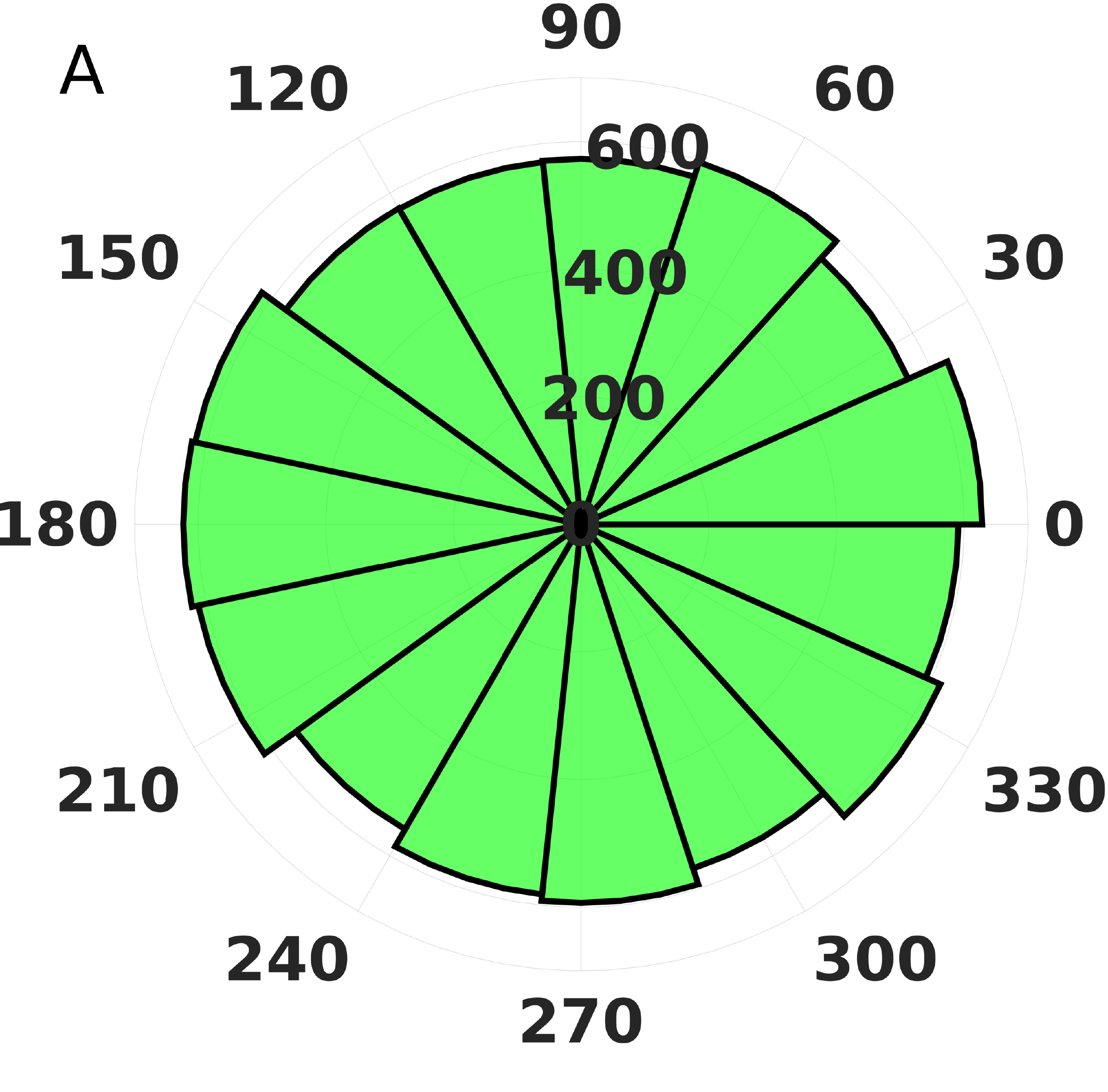}
\includegraphics[scale=0.16]{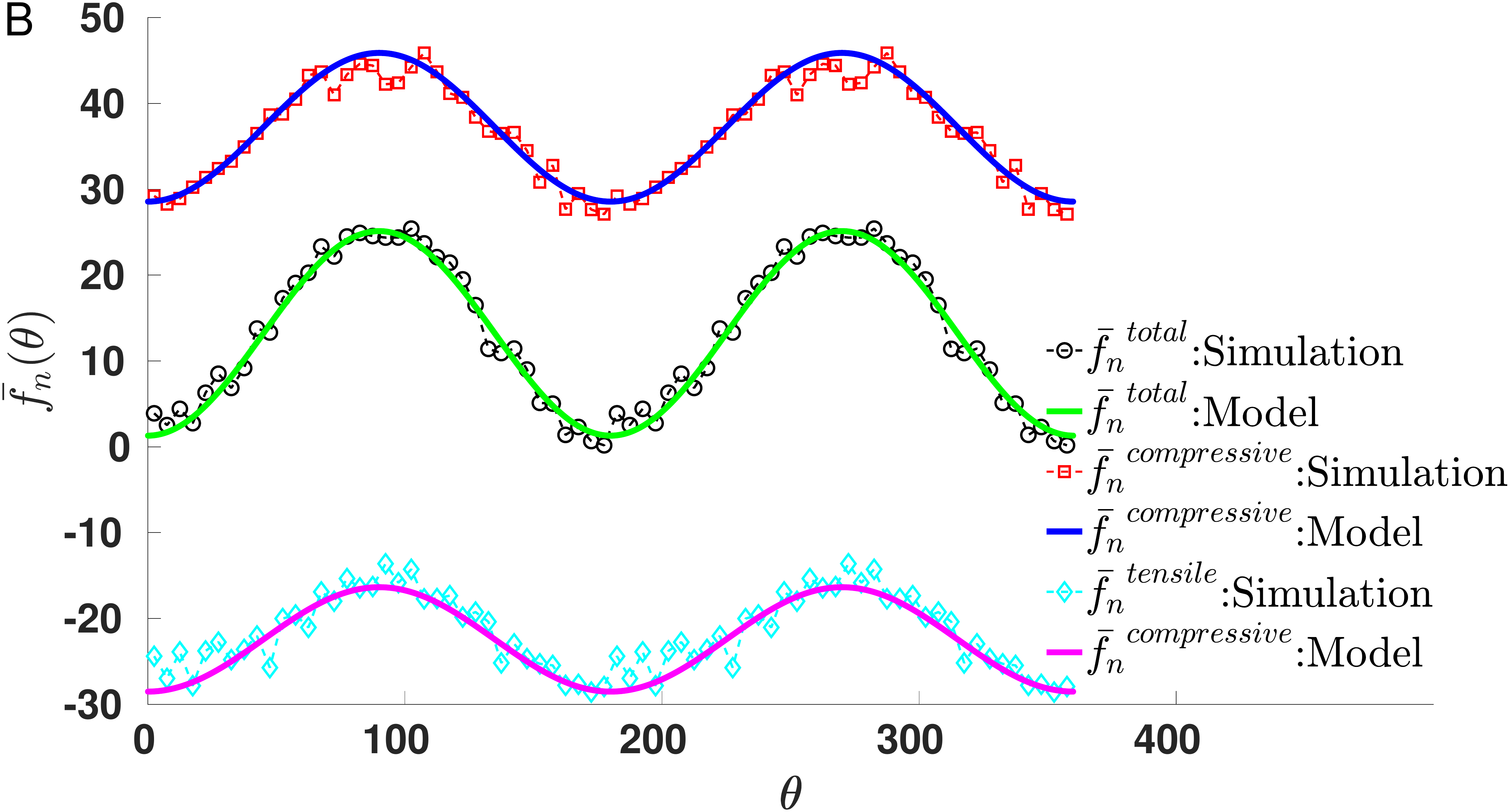}
\includegraphics[scale=0.16]{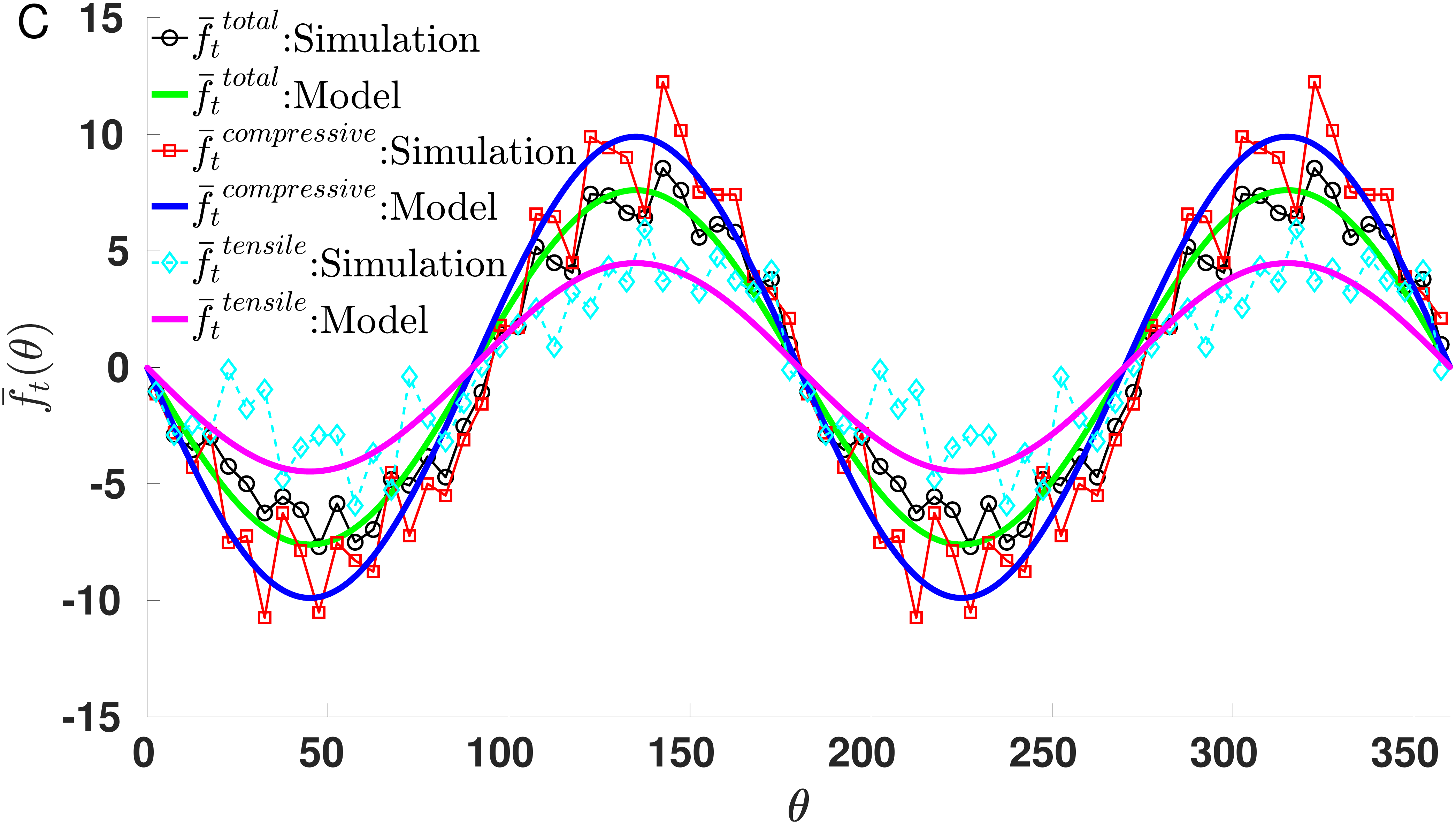}
\includegraphics[scale=0.13]{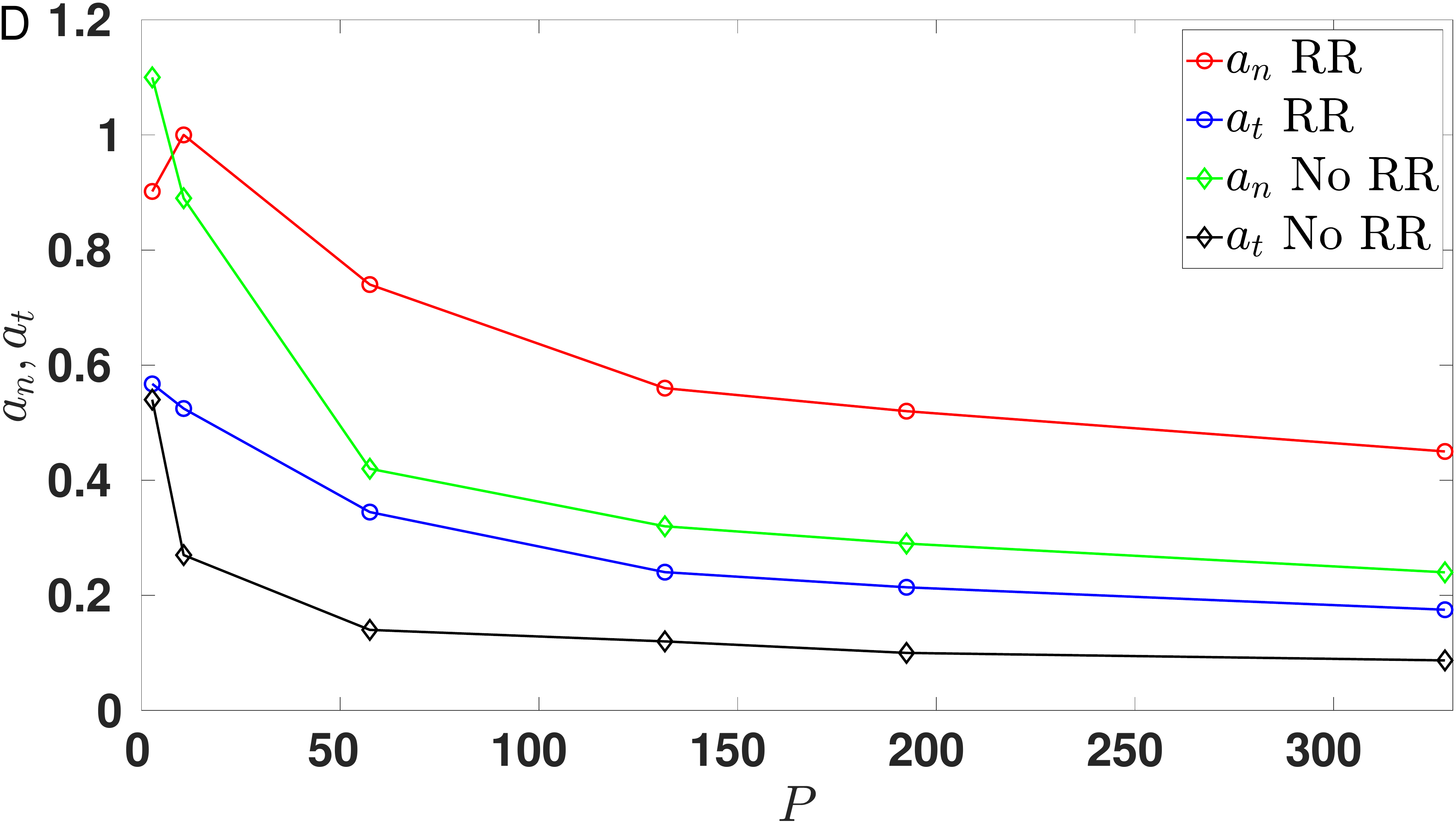}
\includegraphics[scale=0.13]{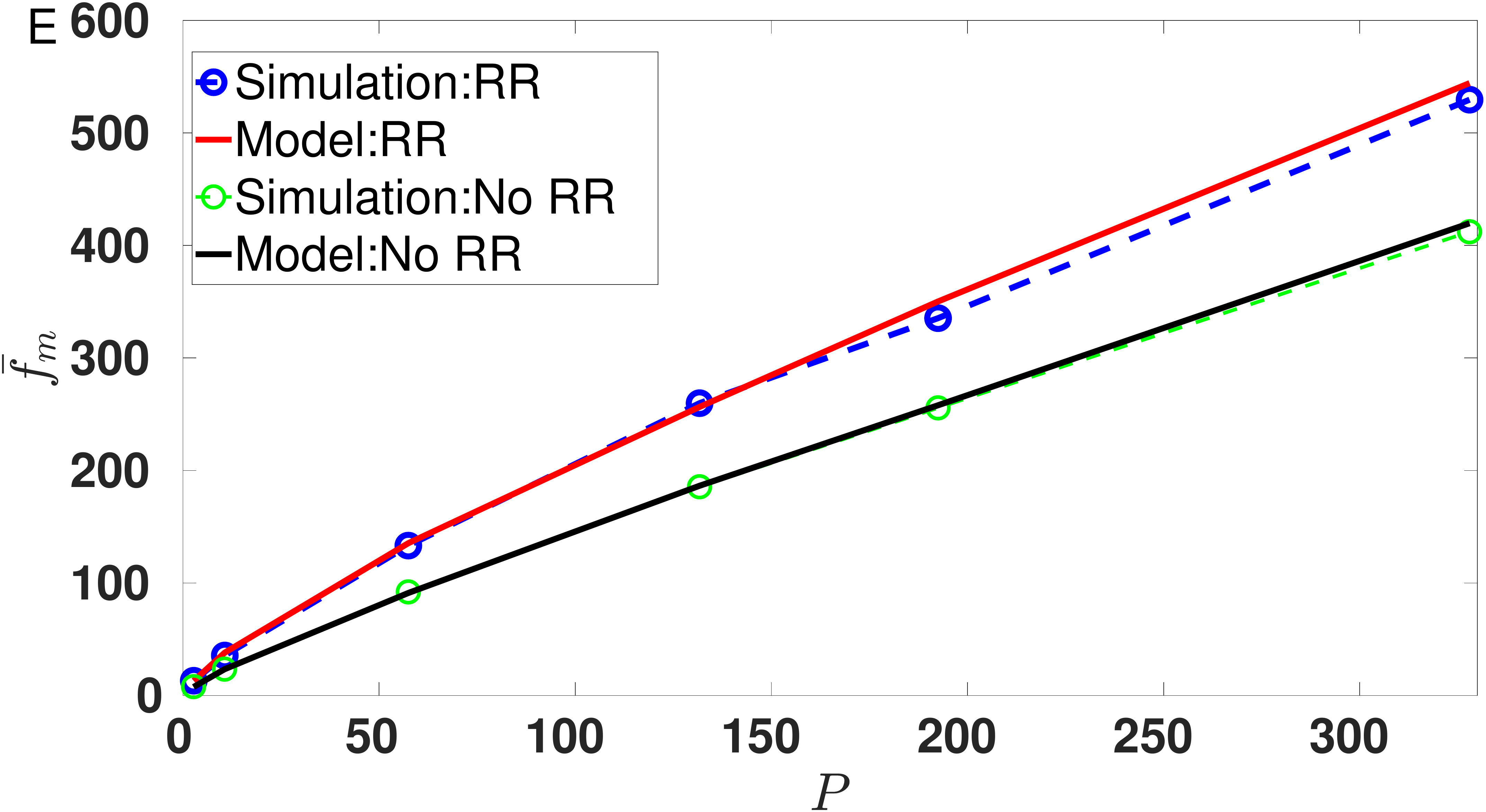}
\includegraphics[scale=0.13]{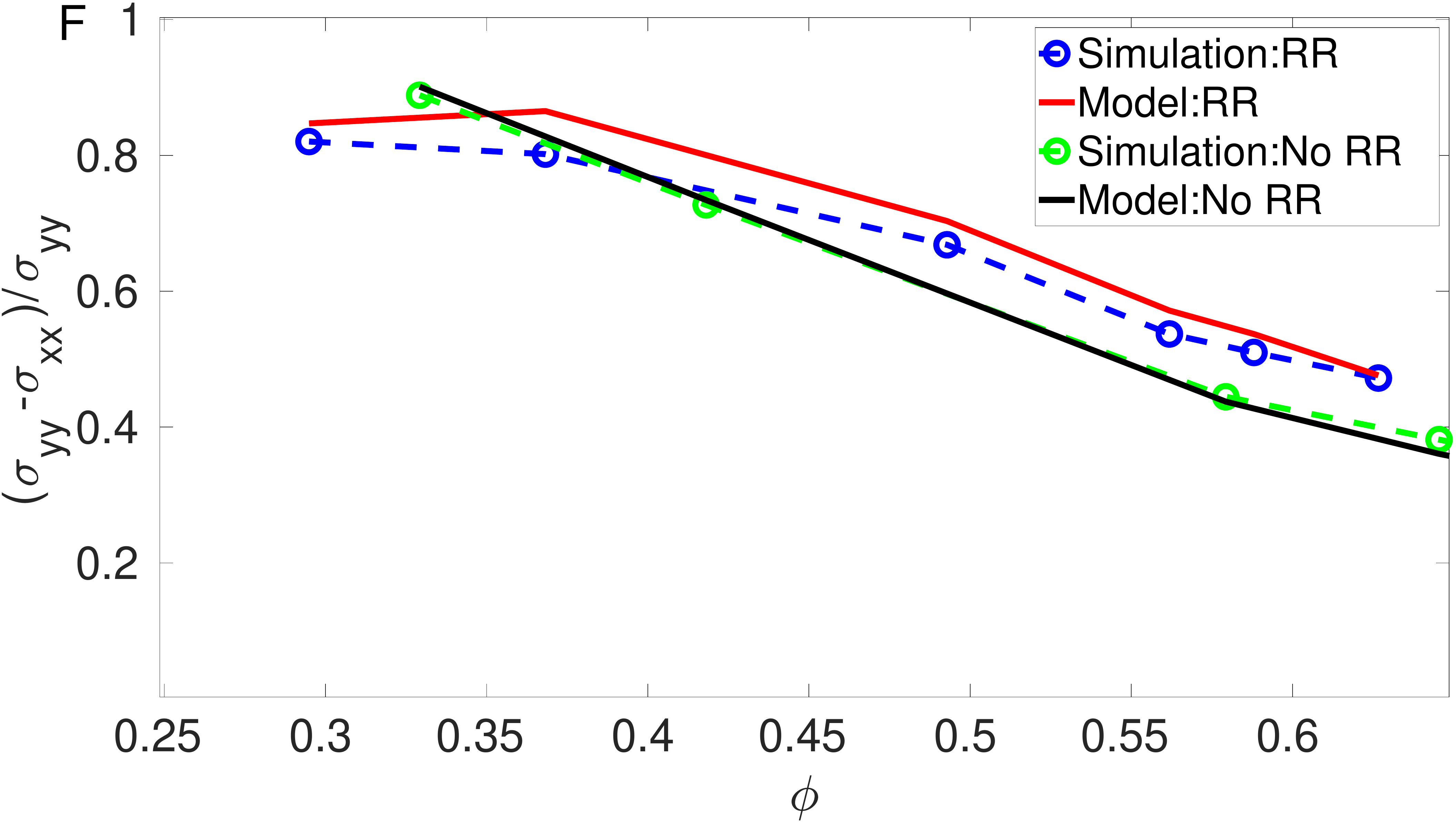} 
\end{center}

\caption{A. Polar histogram of the isotropic contact orientations for $P=2.75$, $V/V_0=18.6$ in the presence of rolling resistance(RR) . Network isotropy is also maintained at high pressure, which is not shown. B. Angular distribution of the normal forces calculated for the total network as well as separately for the tensile and compressive network. Solid curve represents fit to Eq.\ref{fn}. C. Angular distribution for the tangential forces. Solid curve corresponds to the fit to Eq.\ref{ft}. D. Variation of mechanical anisotropy, $a_n$ and $a_t$ as a function of pressure($(\sigma_{xx}+\sigma_{yy})/2$, compression is along $y$ axis). E. Comparison between model prediction and observed mean normal force,$\bar{f}_m$ variation as a function of the system pressure. F. Normal stress difference normalized by the axial stress obtained from the simulation is compared with the model prediction of Eq.\ref{nsd2} for RR and No RR case. Here, $V/V_0=18.6$.}
\label{fig:anisotropy}
\end{figure*}
\begin{figure*}[htbp]
\begin{center}
\includegraphics[scale=0.17]{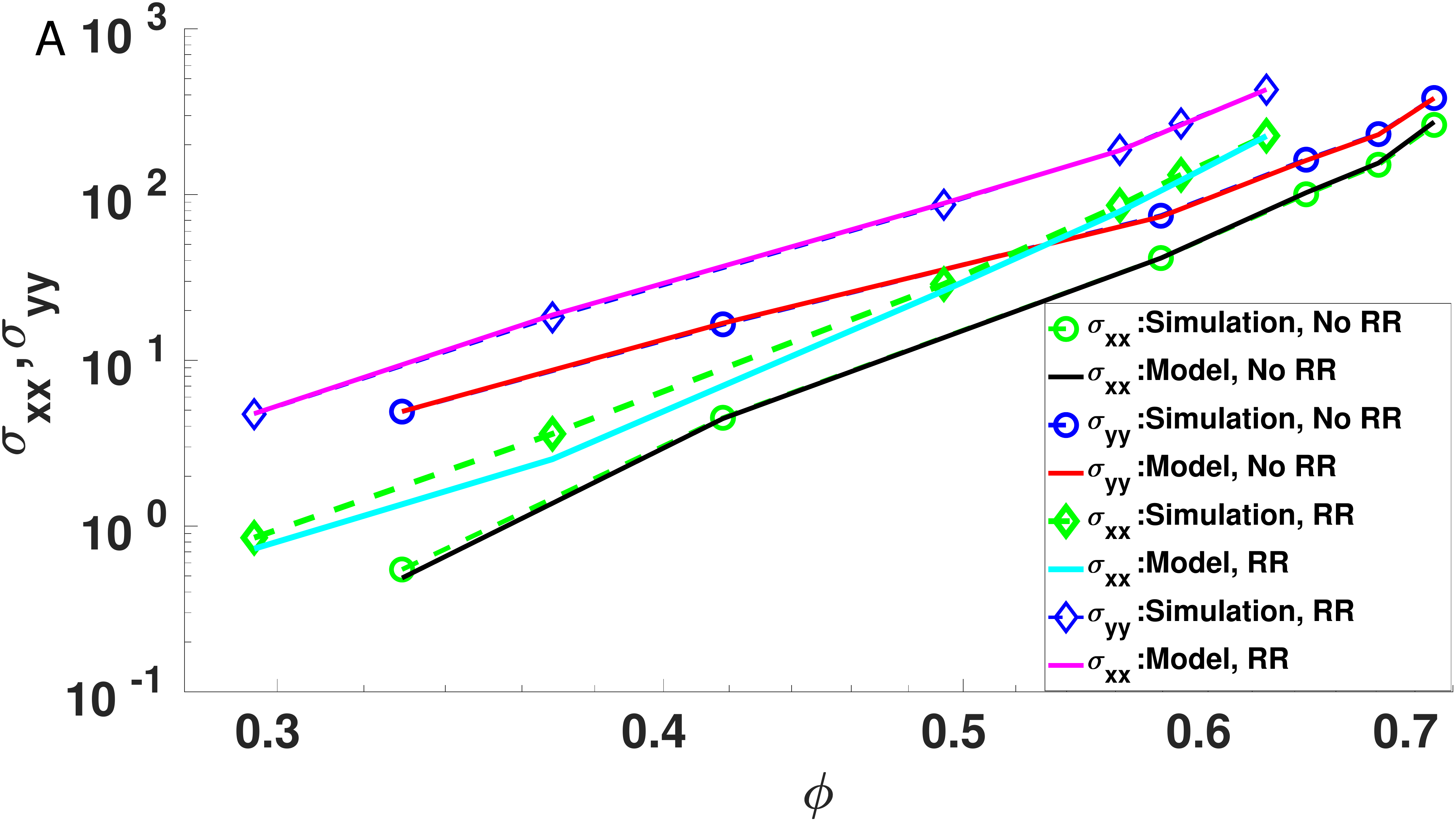}
\includegraphics[scale=0.23]{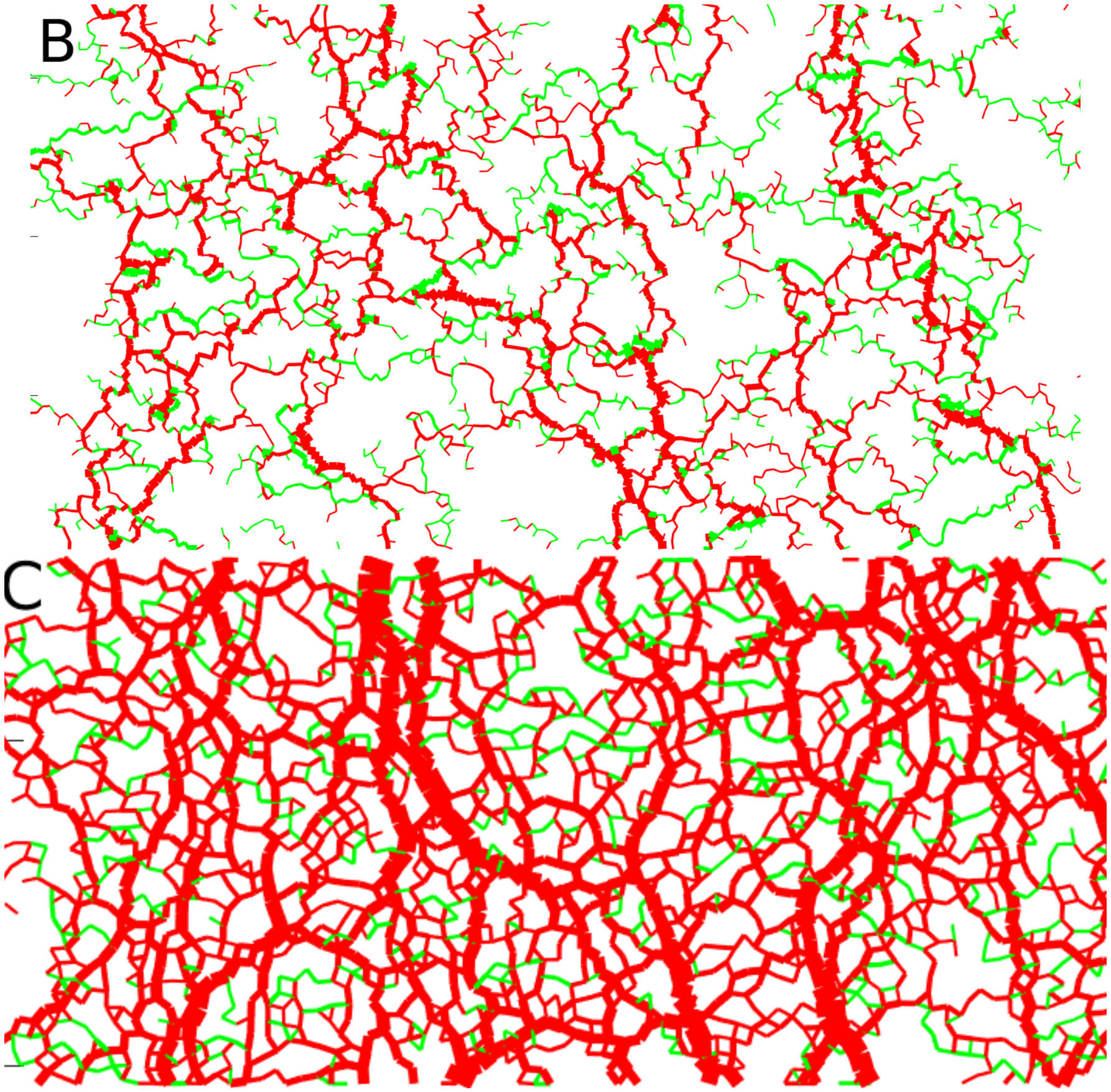}
\includegraphics[scale=0.18]{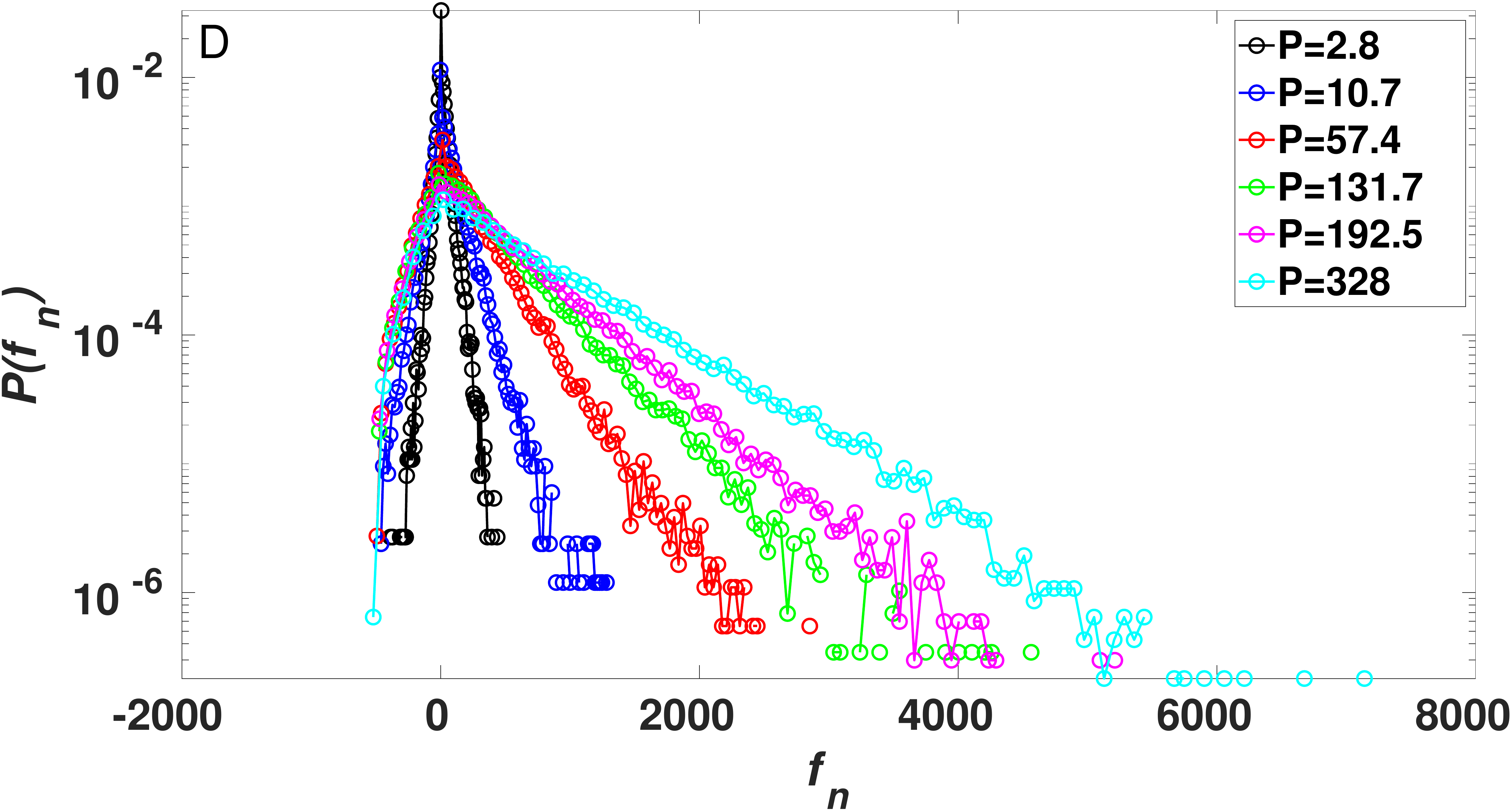}
\includegraphics[scale=0.18]{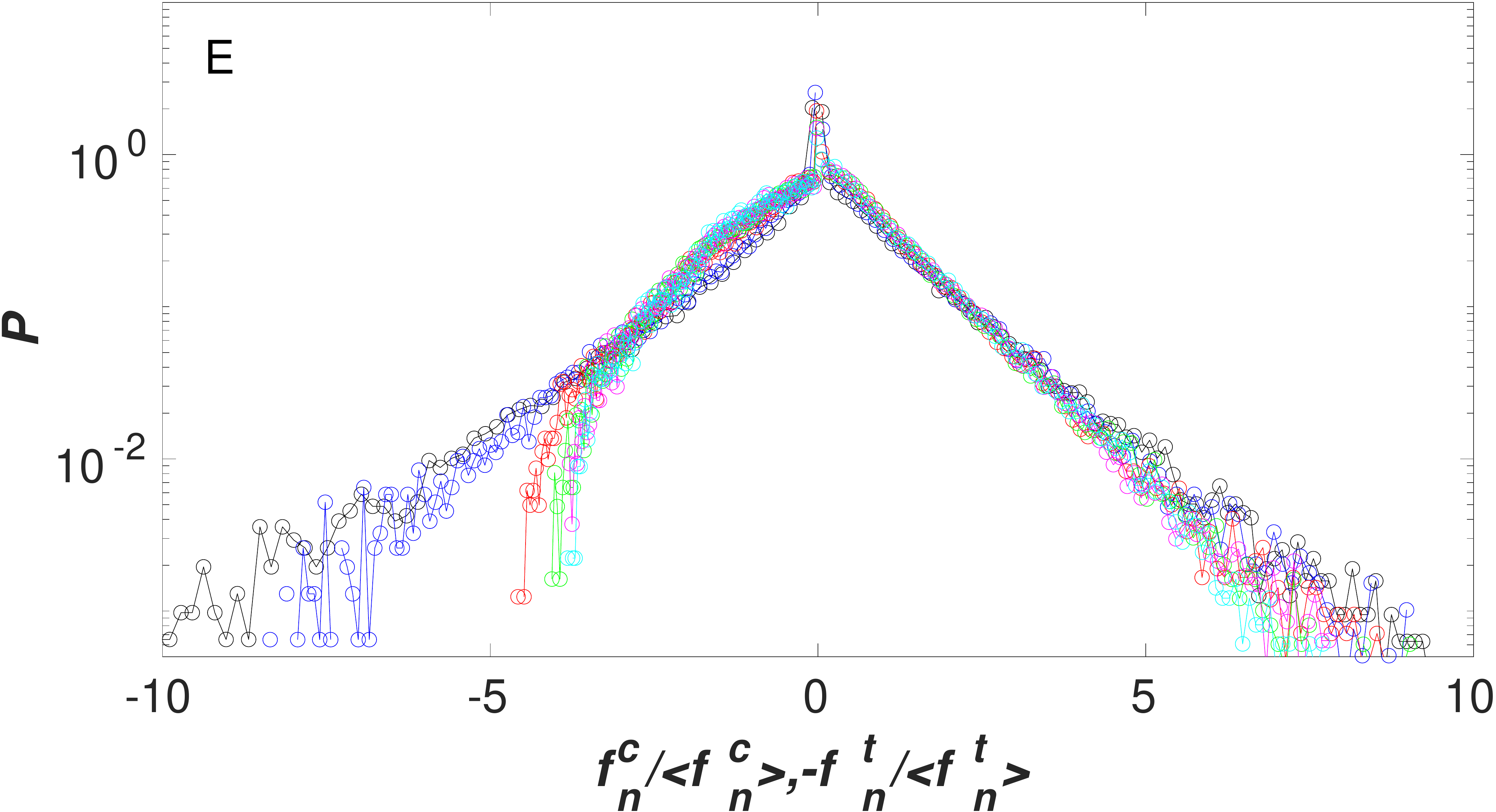}
\end{center}
\caption{A. Comparison between the prediction of the proposed model and the simulation data for the axial, $\sigma_{yy}$ and transverse stress, $\sigma_{xx}$ both in the presence and absence of the rolling resistance(RR). B. Force chain networks in colloidal gel sample very close to the gel point, $P = 2.75$. Here $V/V_0=18.6$ and both the sliding and rolling resistance are present. The compressive force is drawn in red color whereas the green color represents the tensile force. We compute the magnitude of the pairwise normal force and scale the line (joining two centers for a pairwise interaction)thickness as per the magnitude. C. The force chain network far away from the gel point, $P=328$. D. Normal force distribution for the whole force network. E. The tensile ($f_n^{t}$) and compressive normal forces ($f_n^c$) are considered separately and the corresponding distribution is plotted after normalizing the forces by their respective mean($\langle f_n^c \rangle$,$\langle f_n^t \rangle$).}
\label{fig:sigmaxxyy}
\end{figure*}
We begin our analysis by measuring the contact network anisotropy, $a$ and find the network to be almost isotropic ($a=0.02\pm 0.01$) throughout the deformation as evident from the polar histogram of the contact orientations of Fig. \ref{fig:anisotropy}A. The isotropy of the contact network is expected at low pressure (close to the gel point) from the random nature of thermal fluctuations under which the gel is prepared. At high pressure, the isotropy is not apparent due to the anisotropic  uniaxial compression, which suggest very little rearrangement of the contact network due to the attractive interactions. We next examine the anisotropy of the force networks which also play an important role in the stress transmission. Fig. \ref{fig:anisotropy}B shows the angular distribution of the normal force for the total force network. Interestingly, the normal force network is highly anisotropic ($a_n=0.92\pm0.1$) at low pressure, $P=2.75$ with its preferred direction parallel to the axis of compression ($\theta_f=\pi/2$) and the observed distribution matches quite nicely with the approximate form of Eq.\ref{fn}. To investigate the competing effects of normal forces of different sign, we have also plotted the same distribution for the compressive and tensile network separately on top of the total network in Fig. \ref{fig:anisotropy}B. Close inspection reveals that the preferred direction of the anisotropy for both the tensile and compressive network is the same, $\theta_f=\pi/2$, which means the angular distribution of the total normal force network is set by the simple algebraic addition of compressive and tensile part. Fig. \ref{fig:anisotropy}C depicts the angular distribution of the tangential part of the contact force at low pressure and the data suggests significant anisotropy ($a_t=0.57\pm0.06$) in the tangential force network as well.\\

It is also desirable to determine the evolution of the mechanical anisotropy parameters, $a_n,a_t$ as a function of pressure for the correct prediction of the macroscopic stress components as these state parameters encode the history of the deformation. In Fig. \ref{fig:anisotropy}D, $a_n$ and $a_t$ are plotted as a function of the system pressure for cases with and without RR (rolling resistance). In both the cases, mechanical anisotropy parameters decrease with the increasing pressure as the initial anisotropic force chain networks along the axial load direction buckles and transfers part of the axial load to the transverse direction and this transfer of load is resisted by the rolling resistance leading to slow decay of the anisotropy whereas it decays faster in the absence of RR.
\section{Normal stress difference} For the isotropic contact network and noting that $n_c=(4z\phi)/(\pi D^2)$, where $z$ is the average coordination number and $\phi$ is packing fraction,  the system pressure from Eq. \ref{stress} is given as, $P=(n_cD\bar{f}_m)/4=(z \phi \bar{f}_m)/(\pi D)$. This expression of the pressure explicitly assumes no correlations between the contact force and the contact orientations. The predicted relationship between the system pressure and the mean normal force holds almost exactly for all the cases(with and without RR, different thermal energy, different particle size distribution) in our simulation, thus justifying the assumption of no correlations (See Fig. \ref{fig:anisotropy}E). The denominator in the expression for the system pressure contains particle diameter, so another implication of the excellent match between the simulation data and the model is that the particle length scale is the relevant length scale not any fractal length scale as loosely assumed in the existing literature.  Similarly, the expression for the axial ($\sigma_{yy}$), transverse stress ($\sigma_{xx}$) components and subsequently the normalized normal stress difference (NSD) can easily be obtained by substituting Eqns. \ref{Ptheta}, \ref{fn} and \ref{ft} into Eq. \ref{stress} and performing the integration noting that $a\approx0$, $\theta_f=\theta_t=\pi/2$:

\begin{align}
&\sigma_{xx} = P\left(1-\frac{a_n+a_t}{2}\right)\label{sigmaxx2}&\\
&\sigma_{yy} = P\left(1+\frac{a_n+a_t}{2}\right)\label{sigmayy2}&\\
&NSD=\left(\sigma_{yy}-\sigma_{xx}\right)/\sigma_{yy} =\frac{ \left(a_n+a_t\right)}{\left(1+\frac{a_n+a_t}{2}\right)}\label{nsd2}&
\end{align}

The simple form of the expression for the stress components is appealing and it implies that the system pressure and the force anisotropies are the crucial state parameters in the rheology of gel systems. For the isotropic force network (observed in the isotropic compression), axial stress and the transverse stress becomes of the order of system pressure whereas the large anisotropy in the force networks gives rise to large normal stress difference, $\left(\sigma_{yy}-\sigma_{xx}\right)$. The calculation of the system pressure requires the knowledge of the mean normal force, $\bar{f}_m$, average coordination number, $z$ and the packing fraction, $\phi$. Unlike the frictionless granular matter, there is no unique relationship between $z$ and $\phi$ for the gel system formed at low packing fractions and the variation of $z$ as a function of $\phi$ strongly depends on the preparation protocol and the interaction potential. In the absence of RR (only sliding friction is present), $z$ almost remains constant at $3.4$ with the change in $\phi$ from $0.3$ to $0.7$ whereas $z$ changes from $2$ to $3$ in the presence of RR. The variation of the mean normal force follows a power law scaling with the packing fraction and strongly dictates the variation of the system pressure which is also a power law with $\phi$. \\
In summary, $z$, $\bar{f}_m$, $a_n$ and $a_t$ are the primary internal state parameters whose evolution with $\phi$ will determine the evolution of stress components as a function of $\phi$. In Fig. \ref{fig:anisotropy}F, we show the comparison between the model prediction (Eq. \ref{nsd2}) and the observed normal stress difference normalized by the axial stress in the presence of RR and find excellent agreement. The nice match between the model and the simulation data is also observed in the absence of RR, thus pointing towards the robustness of the model. It is important to note that the expression for the NSD does not contain $z$, $\phi$ and $\bar{f}_m$ and the evolution of the NSD can successfully be determined only by tracking the evolution of the force anisotropy parameters accurately. In essence, anisotropy in the force networks lead to the normal stress difference in the soft solid like amorphous materials. In order to gain further confidence in the proposed model, we also compare the model prediction for the axial stress, $\sigma_{xx}$ and the transverse stress, $\sigma_{yy}$ with that of the simulation data in Fig. \ref{fig:sigmaxxyy}A and the agreement is excellent. For the comparison, the values of the primary internal state parameters measured in the simulation are fed to the model. The proposed equations (Eq.\ref{sigmaxx2}-Eq.\ref{nsd2}) relating the different components of the stress tensor to the  appropriate internal state parameters,  can be considered as the constitutive relations for the gel system as it correctly describes the state of the stress based on other discernible characteristics of its physical state. Most importantly, the effect of construction history is adequately captured via the identified crucial state parameters, which is otherwise impossible in the traditional constitutive relation which relates stress to the strain irrespective of the preparation history and quenched stresses. Next, to elucidate the origin of the non linear growth of stress with packing fraction, we turn our attention to the normal force distribution as the exponent of the power law growth of stress is strongly influenced by the variation of the mean normal force, $\bar{f}_m$ with $\phi$.
\begin{figure*}[htbp]
\begin{center}
\includegraphics[scale=0.2]{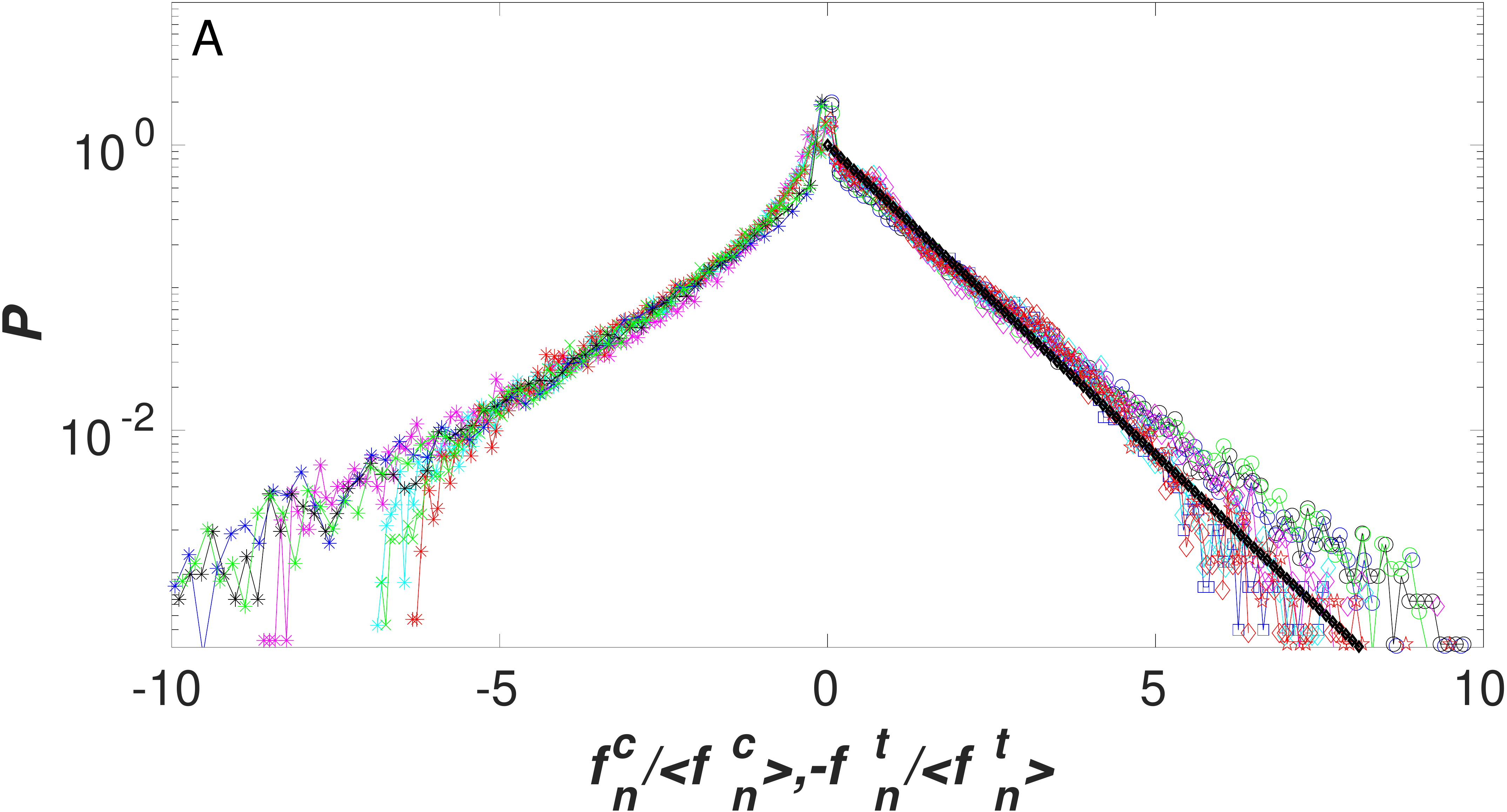}
\includegraphics[scale=0.2]{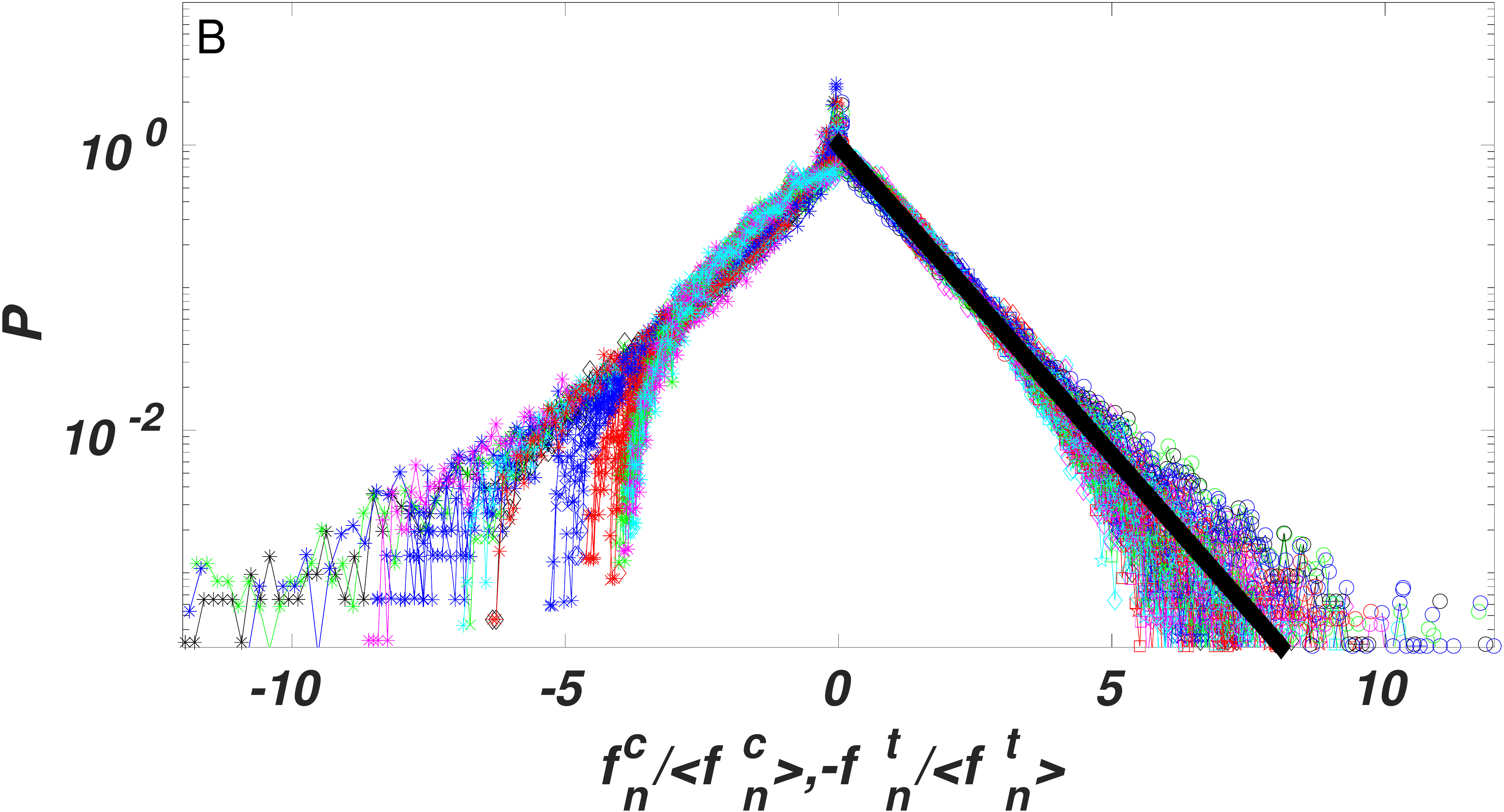}
\end{center}
\caption{A. Universal normal force distribution close to the gel point. The distribution for the tensile and compressive normal forces are plotted separately after normalizing the forces by their respective mean. Note that the tensile force is cut off by the attraction force, $F_0$. In this plot, we used different values of $V/V_0$ ($5.9,18.6$ and $58.7$), $F_0/k_nD$ ($0.0025$ and $0.0005$), particle size distribution (monodisperse and bidisperse), different contact models (Hookean and Hertzian), with and without the rolling resistance and observe a nice collapse for the large parametric space. The solid black line represents the maximum entropy prediction, $e^{-f_n^{c}/<f_n^c>}$ for the compressive part. B. The same collapse at high pressure where large deviations in the tails of the compressive and the tensile part are observed.}
\label{fig:collapse}
\end{figure*}
\section{Universal force distribution} Stresses in soft amorphous materials get transmitted in a highly heterogeneous fashion via selected pathways known as force chains. Extensive numerical and experimental investigations \cite{liu1995force,mueth1998force,majmudar2005contact,radjai1996force,o2002random,akella2018force,radjai1998bimodal} in dry granular matter revealed several interesting aspects of the force chain networks and the corresponding force distribution, such as organization of the force chain network into weak and strong network, exponential distribution for the strong forces and nearly uniform or power law distribution for the small forces. Although a voluminous amount of studies exist for the dry granular materials and a somewhat universal picture of the stress transmission emerges, very few investigations\cite{richefeu2009force,richefeu2006stress,richefeu2006shear,gilabert2007computer} were directed towards cohesive granular matter and colloidal gel systems jammed at very low packing fractions. In Fig. \ref{fig:sigmaxxyy}B-C, we present the typical normal force chain networks formed in the colloidal gel systems at low and high pressure. Contrary to the dry granular materials, the tensile normal force is admissible in the gel materials due to the attractive interactions. The force chain network close to the gel point clearly consists of two complimentary percolating networks of compressive and tensile forces with large portions where these two networks intermingle. For the compressive network, strong forces form the backbone of the stress transmission. As the system pressure increases, the tensile force network no longer percolates the whole system rather it gets localized inside the strongly percolating compressive force network.\\
To obtain a quantitative picture of the force networks, we plot the normal force distribution in Fig. \ref{fig:sigmaxxyy}D as a function of the system pressure. The distribution is initially symmetric about zero at low pressure and becomes asymmetric with broader compressive part as the system pressure increases. Close scrutiny of the data reveals a power law divergence for the small forces close to the gel point, which is not present at high pressure although a peak close to the zero force exists. The presence of a peak at $f_n\approx0$ even at higher pressure clearly suggests a non-trivial stress transmission in the colloidal gel systems. Zero normal force corresponds to those contacts where the attraction is balanced by the elastic repulsion, clearly these contacts are stress transmitting unlike the usual rattlers in dry granular matter. The abundance of such contacts implies that the fraction of such contacts can easily become a part of the new force chain network whenever an incompatible load is applied. The compressive part of the distribution follows an exponential distribution for all pressures and the distribution for all pressures does not collapse when $f_n$ is normalized by the system mean normal force,$<f_n>$ or the strength of attraction. \\
The existence of two separate percolating force networks at low pressure motivates us to consider the tensile and compressive part of the normal forces separately and normalizing the forces of each part by their respective mean lead to the collapse of the distribution for the tensile and compressive part for all pressures (Fig. \ref{fig:sigmaxxyy}E). The tensile normal force distribution is exponential initially at low pressure and is a mirror image of the compressive part, and latter on the symmetry is lost and the distribution is no longer exponential. The nice collapse for the compressive part implies a persistent percolating compressive network which grows stronger with the system pressure and the distribution remains exponential throughout the deformation with the mean compressive normal force scaling with the system pressure.  All the essential parameters are now changed to mimic different gel preparation history, gel microstructure, particle interaction potential, particle size dispersity and Fig. \ref{fig:collapse}A depicts a remarkable collapse of the normal force distribution close to the gel point when separately plotted for the compressive and tensile part after normalizing by the respective mean normal force. The collapse of the data is observed for a huge parameter space as mentioned in the figure caption. The distribution of the compressive part almost exactly follows the maximum entropy prediction ($e^{-f_n^{c}/<f_n^c>}$) \cite{akella2018force} of the distribution subjected to the constraint, $f_n^{c}/<f_n^c>\, =1$ whereas the tensile part is exponential with a slower decay rate than the compressive part. The exponential distribution of the forces is also hypothesized from the experimental observations \cite{dinsmore2002direct} of exponential distribution of spring constants. Fig. \ref{fig:collapse}B shows the collapse considering the high pressure data for different parameters, although the collapse is reasonable but large deviations in the tails of the compressive and the tensile part are observed. The universality in the force distribution at low pressure points towards a non-trivial self organization of the force chain networks into two coexisting percolating compressive and tensile force networks that support the external load. Just at the gel point, under no confining stresses, the force distribution of the normal forces of opposite sign has to be exactly symmetric about zero so that the mean normal force becomes zero in line with the condition of zero confining stress. It is very important to note that the zero confining stress does not imply zero interaction force at all contacts i.e the perfect balance between elastic repulsion and attraction which is rarely established in the conventional gel preparation protocol. The equilibrium static structure at the gel point can contain large compressive and tensile forces, and a delicate balance between normal forces of opposite sign make the mean normal force zero. As the system confining pressure increases, the mean normal force becomes positive and the scale is set by the competition between compressive and tensile force distributions. Surprisingly it is further observed that the tensile and compressive normal forces that are above the system's mean normal force contribute almost $95\%$ to the mean normal force. So in summary, the subtle interplay between the tails of the distribution for the compressive and tensile parts dictates the magnitude of the mean normal force in the gel materials. 
\begin{figure*}[htbp]
\begin{center}
\includegraphics[scale=0.2]{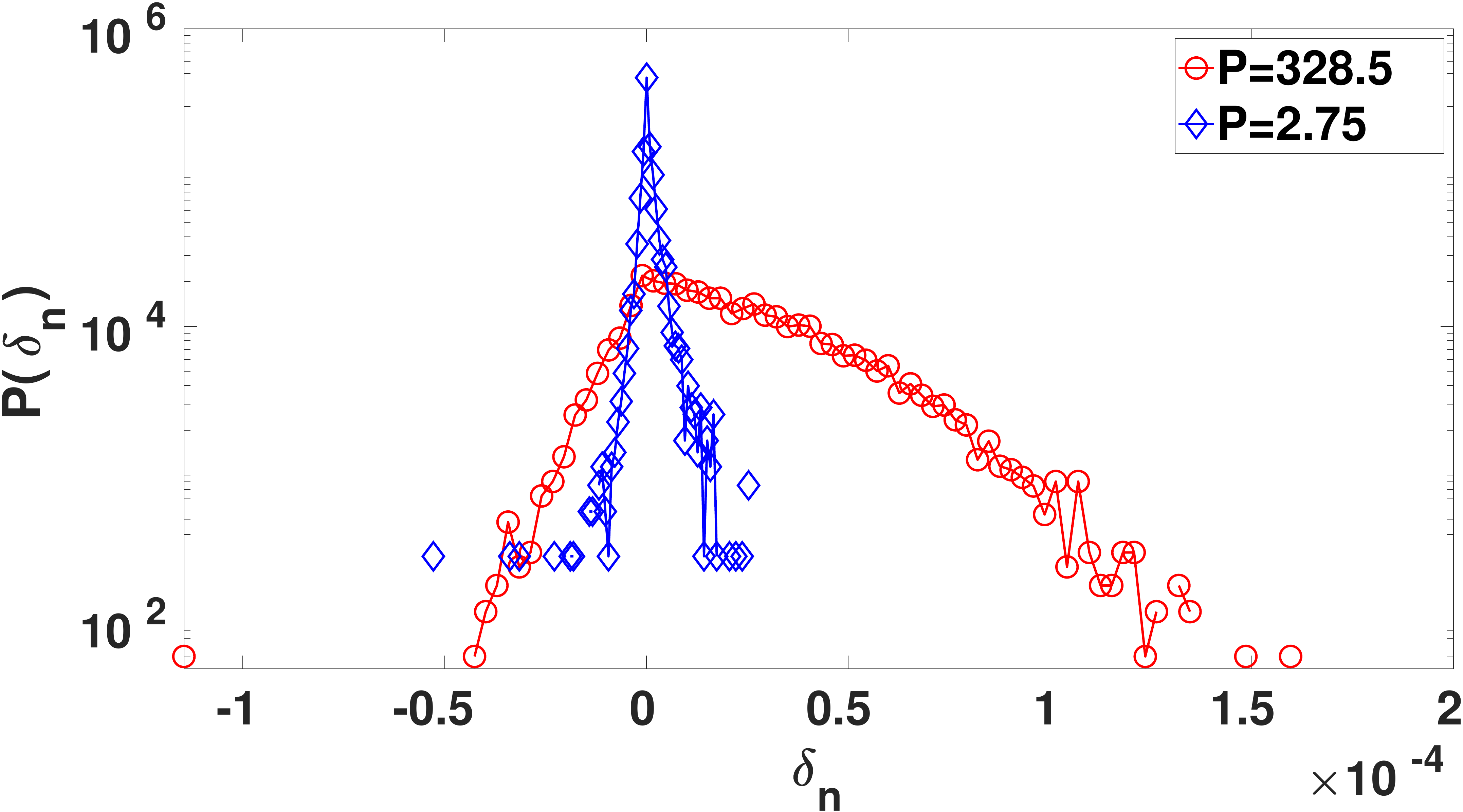}
\end{center}
\caption{Probability distribution of the normal components of the displacements for low and high pressure. Here $V/V_0 = 58.7$, $F_0/k_nD= 0.0025$ and both the rolling and sliding friction is present.}
\label{fig:smallstrain}
\end{figure*}
\section{Small strain response} In order to gain further insights into the nonlinear growth of the mean normal force with $\phi$, we also investigated the distribution of the displacements parallel to the line connecting centers of the interacting particles under small strain ($\sim 1\times10^{-4}$). The mean normal force is expected to scale as $\bar{f}_m \sim \langle\delta_n\rangle$ and the variation of the mean normal deformation as a function of the packing fraction should reflect the scaling of the mean normal force with $\phi$. The non-affinity in the displacement field not only arises due to the tangential motions in the form of sliding and rolling, but it can also arise due to the stretching of the bonds under a global compressive strain. Close to the gel point($P=2.75$), the probability distribution of the normal displacements is almost symmetric about zero (See Fig. \ref{fig:smallstrain}) leading to very small mean normal deformation due to the competing effects between positive (compression) and negative displacements (extension). The symmetry is lost at high pressure and the compression deformation dominates the stretching giving rise to large mean normal deformation. In all cases, a peak around zero is observed signifying the presence of large proportion of contacts that stores no elastic energy. The power law exponent for the growth of the mean normal deformation with $\phi$ is similar to that observed for the variation of mean normal force with $\phi$.
\section{Discussion and conclusion}
In this work, we perform extensive numerical simulations to investigate the different aspects of the stress transmission in strong colloidal gel system (strength of attraction, $U$ is very large compared to the thermal energy, $k_BT$) subjected to uniaxial compression, which is important in diverse industrial process such as film formation in paints and coatings, solid liquid separation process and fabrication of ceramic materials. Our simulation methodology presents a novel approach to study the colloidal gel systems taking into account sliding and rolling resistance that are present due to the nanoscale roughness. Inspired by the granular literature \cite{rothenburg1989analytical,radjai1998bimodal}, we develop a relationship between the stress tensor and different crucial state parameters of the system such as mean normal force, anisotropy of the inhomogenous force networks and find excellent agreement with the simulation results. The normal stress difference is now correctly predicted after accounting for the mechanical anisotropy in the normal and tangential forces. 

We also observe that the mean normal force of the system follows a simple relationship with the system pressure and dictates the scale of the stress in the system. The distribution of the tensile and compressive normal forces show nice collapse when normalized by their respective mean, which point towards the existence of two coexisting percolating force networks: tensile and compressive. Close to the gel point, the distribution of the tensile and compressive normal forces is exponential and symmetric about zero, and when normalized by their respective mean show universal behaviour irrespective of different interaction potential, particle size distribution and thermal energy. At high pressure, the collapse of the distribution is also observed but with deviations in the tails and asymmetry about zero. The maximum entropy prediction of the distribution for the compressive forces subjected to the mean normal force for the compressive part leads to a nice fit to the observed distribution. Interestingly, tensile and compressive normal forces that are above the system's mean normal force contribute almost $95\%$ to the mean normal force. The delicate balance between the tail of the distributions for compressive and tensile normal forces set the scale of mean normal force in the system. We also perform simulations to study the small strain response of the gel at different pressures and find that the distribution of the normal displacements (extension or compression) at the contact is exponential and symmetric about zero at low pressure and the symmetry is lost at high pressure.

Our results strongly indicate that the competition between the extension and compression at the particle length scale decides the mean normal deformation of the system and thus the mean normal force. Due to the symmetry close to the gel point, cancellation effects of positive and negative displacements is the strongest leading to almost negligible mean normal deformation. The observed power law growth of stress with packing fraction in the simulation is directly linked to the power law growth of the mean force, which is effected by the subtle interplay between extension and compression deformations at the particle length scale. Overall, our results show that the elastic response of the gel network is controlled by the particle length scale phenomenon with a complex interplay between compression and tension at the particle contact. This work provides a novel microscopic viewpoint to understand the gel's mechanical response and future work should focus on unraveling the role of the contacts with zero normal force as the peak around $f_n\approx0$ clearly suggests some sort of nontrivial self organization of the network to support the load.
\textit{Acknowledgement-}S.R. acknowledges the support of the Science and Engineering Research Board, DST, India under grant no. SRG/2020/001943 and the IIT Ropar under ISIRD grant..

\bibliography{ref2}

\begin{thebibliography}{40}%
\makeatletter
\providecommand \@ifxundefined [1]{%
 \@ifx{#1\undefined}
}%
\providecommand \@ifnum [1]{%
 \ifnum #1\expandafter \@firstoftwo
 \else \expandafter \@secondoftwo
 \fi
}%
\providecommand \@ifx [1]{%
 \ifx #1\expandafter \@firstoftwo
 \else \expandafter \@secondoftwo
 \fi
}%
\providecommand \natexlab [1]{#1}%
\providecommand \enquote  [1]{``#1''}%
\providecommand \bibnamefont  [1]{#1}%
\providecommand \bibfnamefont [1]{#1}%
\providecommand \citenamefont [1]{#1}%
\providecommand \href@noop [0]{\@secondoftwo}%
\providecommand \href [0]{\begingroup \@sanitize@url \@href}%
\providecommand \@href[1]{\@@startlink{#1}\@@href}%
\providecommand \@@href[1]{\endgroup#1\@@endlink}%
\providecommand \@sanitize@url [0]{\catcode `\\12\catcode `\$12\catcode
  `\&12\catcode `\#12\catcode `\^12\catcode `\_12\catcode `\%12\relax}%
\providecommand \@@startlink[1]{}%
\providecommand \@@endlink[0]{}%
\providecommand \url  [0]{\begingroup\@sanitize@url \@url }%
\providecommand \@url [1]{\endgroup\@href {#1}{\urlprefix }}%
\providecommand \urlprefix  [0]{URL }%
\providecommand \Eprint [0]{\href }%
\providecommand \doibase [0]{http://dx.doi.org/}%
\providecommand \selectlanguage [0]{\@gobble}%
\providecommand \bibinfo  [0]{\@secondoftwo}%
\providecommand \bibfield  [0]{\@secondoftwo}%
\providecommand \translation [1]{[#1]}%
\providecommand \BibitemOpen [0]{}%
\providecommand \bibitemStop [0]{}%
\providecommand \bibitemNoStop [0]{.\EOS\space}%
\providecommand \EOS [0]{\spacefactor3000\relax}%
\providecommand \BibitemShut  [1]{\csname bibitem#1\endcsname}%
\let\auto@bib@innerbib\@empty
\bibitem [{\citenamefont {Zaccarelli}(2007)}]{zaccarelli2007colloidal}%
  \BibitemOpen
  \bibfield  {author} {\bibinfo {author} {\bibfnamefont {E.}~\bibnamefont
  {Zaccarelli}},\ }\href@noop {} {\bibfield  {journal} {\bibinfo  {journal}
  {Journal of Physics: Condensed Matter}\ }\textbf {\bibinfo {volume} {19}},\
  \bibinfo {pages} {323101} (\bibinfo {year} {2007})}\BibitemShut {NoStop}%
\bibitem [{\citenamefont {Dagur}\ \emph {et~al.}(2022)\citenamefont {Dagur},
  \citenamefont {Mondal},\ and\ \citenamefont {Roy}}]{dagur2022spatial}%
  \BibitemOpen
  \bibfield  {author} {\bibinfo {author} {\bibfnamefont {D.~S.}\ \bibnamefont
  {Dagur}}, \bibinfo {author} {\bibfnamefont {C.}~\bibnamefont {Mondal}}, \
  and\ \bibinfo {author} {\bibfnamefont {S.}~\bibnamefont {Roy}},\ }\href@noop
  {} {\bibfield  {journal} {\bibinfo  {journal} {arXiv preprint
  arXiv:2205.11575}\ } (\bibinfo {year} {2022})}\BibitemShut {NoStop}%
\bibitem [{\citenamefont {Hsiao}\ \emph {et~al.}(2012)\citenamefont {Hsiao},
  \citenamefont {Newman}, \citenamefont {Glotzer},\ and\ \citenamefont
  {Solomon}}]{hsiao2012role}%
  \BibitemOpen
  \bibfield  {author} {\bibinfo {author} {\bibfnamefont {L.~C.}\ \bibnamefont
  {Hsiao}}, \bibinfo {author} {\bibfnamefont {R.~S.}\ \bibnamefont {Newman}},
  \bibinfo {author} {\bibfnamefont {S.~C.}\ \bibnamefont {Glotzer}}, \ and\
  \bibinfo {author} {\bibfnamefont {M.~J.}\ \bibnamefont {Solomon}},\
  }\href@noop {} {\bibfield  {journal} {\bibinfo  {journal} {Proceedings of the
  National Academy of Sciences}\ }\textbf {\bibinfo {volume} {109}},\ \bibinfo
  {pages} {16029} (\bibinfo {year} {2012})}\BibitemShut {NoStop}%
\bibitem [{\citenamefont {Tsurusawa}\ \emph {et~al.}(2019)\citenamefont
  {Tsurusawa}, \citenamefont {Leocmach}, \citenamefont {Russo},\ and\
  \citenamefont {Tanaka}}]{tsurusawa2019direct}%
  \BibitemOpen
  \bibfield  {author} {\bibinfo {author} {\bibfnamefont {H.}~\bibnamefont
  {Tsurusawa}}, \bibinfo {author} {\bibfnamefont {M.}~\bibnamefont {Leocmach}},
  \bibinfo {author} {\bibfnamefont {J.}~\bibnamefont {Russo}}, \ and\ \bibinfo
  {author} {\bibfnamefont {H.}~\bibnamefont {Tanaka}},\ }\href@noop {}
  {\bibfield  {journal} {\bibinfo  {journal} {Science advances}\ }\textbf
  {\bibinfo {volume} {5}},\ \bibinfo {pages} {eaav6090} (\bibinfo {year}
  {2019})}\BibitemShut {NoStop}%
\bibitem [{\citenamefont {Pantina}\ and\ \citenamefont
  {Furst}(2005)}]{pantina2005elasticity}%
  \BibitemOpen
  \bibfield  {author} {\bibinfo {author} {\bibfnamefont {J.~P.}\ \bibnamefont
  {Pantina}}\ and\ \bibinfo {author} {\bibfnamefont {E.~M.}\ \bibnamefont
  {Furst}},\ }\href@noop {} {\bibfield  {journal} {\bibinfo  {journal}
  {Physical review letters}\ }\textbf {\bibinfo {volume} {94}},\ \bibinfo
  {pages} {138301} (\bibinfo {year} {2005})}\BibitemShut {NoStop}%
\bibitem [{\citenamefont {Furst}\ and\ \citenamefont
  {Pantina}(2007)}]{furst2007yielding}%
  \BibitemOpen
  \bibfield  {author} {\bibinfo {author} {\bibfnamefont {E.~M.}\ \bibnamefont
  {Furst}}\ and\ \bibinfo {author} {\bibfnamefont {J.~P.}\ \bibnamefont
  {Pantina}},\ }\href@noop {} {\bibfield  {journal} {\bibinfo  {journal}
  {Physical Review E}\ }\textbf {\bibinfo {volume} {75}},\ \bibinfo {pages}
  {050402} (\bibinfo {year} {2007})}\BibitemShut {NoStop}%
\bibitem [{\citenamefont {Kantor}\ and\ \citenamefont
  {Webman}(1984)}]{kantor1984elastic}%
  \BibitemOpen
  \bibfield  {author} {\bibinfo {author} {\bibfnamefont {Y.}~\bibnamefont
  {Kantor}}\ and\ \bibinfo {author} {\bibfnamefont {I.}~\bibnamefont
  {Webman}},\ }\href@noop {} {\bibfield  {journal} {\bibinfo  {journal}
  {Physical Review Letters}\ }\textbf {\bibinfo {volume} {52}},\ \bibinfo
  {pages} {1891} (\bibinfo {year} {1984})}\BibitemShut {NoStop}%
\bibitem [{\citenamefont {Shih}\ \emph {et~al.}(1990)\citenamefont {Shih},
  \citenamefont {Shih}, \citenamefont {Kim}, \citenamefont {Liu},\ and\
  \citenamefont {Aksay}}]{shih1990scaling}%
  \BibitemOpen
  \bibfield  {author} {\bibinfo {author} {\bibfnamefont {W.-H.}\ \bibnamefont
  {Shih}}, \bibinfo {author} {\bibfnamefont {W.~Y.}\ \bibnamefont {Shih}},
  \bibinfo {author} {\bibfnamefont {S.-I.}\ \bibnamefont {Kim}}, \bibinfo
  {author} {\bibfnamefont {J.}~\bibnamefont {Liu}}, \ and\ \bibinfo {author}
  {\bibfnamefont {I.~A.}\ \bibnamefont {Aksay}},\ }\href@noop {} {\bibfield
  {journal} {\bibinfo  {journal} {Physical review A}\ }\textbf {\bibinfo
  {volume} {42}},\ \bibinfo {pages} {4772} (\bibinfo {year}
  {1990})}\BibitemShut {NoStop}%
\bibitem [{\citenamefont {Roy}\ and\ \citenamefont
  {Tirumkudulu}(2016{\natexlab{a}})}]{roy2016b}%
  \BibitemOpen
  \bibfield  {author} {\bibinfo {author} {\bibfnamefont {S.}~\bibnamefont
  {Roy}}\ and\ \bibinfo {author} {\bibfnamefont {M.~S.}\ \bibnamefont
  {Tirumkudulu}},\ }\href@noop {} {\bibfield  {journal} {\bibinfo  {journal}
  {J. Rheol.}\ }\textbf {\bibinfo {volume} {60}},\ \bibinfo {pages} {575}
  (\bibinfo {year} {2016}{\natexlab{a}})}\BibitemShut {NoStop}%
\bibitem [{\citenamefont {Roy}\ and\ \citenamefont
  {Tirumkudulu}(2020)}]{roy2020micro}%
  \BibitemOpen
  \bibfield  {author} {\bibinfo {author} {\bibfnamefont {S.}~\bibnamefont
  {Roy}}\ and\ \bibinfo {author} {\bibfnamefont {M.~S.}\ \bibnamefont
  {Tirumkudulu}},\ }\href@noop {} {\bibfield  {journal} {\bibinfo  {journal}
  {Soft Matter}\ } (\bibinfo {year} {2020})}\BibitemShut {NoStop}%
\bibitem [{\citenamefont {Roy}\ and\ \citenamefont
  {Tirumkudulu}(2016{\natexlab{b}})}]{roy2016universality}%
  \BibitemOpen
  \bibfield  {author} {\bibinfo {author} {\bibfnamefont {S.}~\bibnamefont
  {Roy}}\ and\ \bibinfo {author} {\bibfnamefont {M.~S.}\ \bibnamefont
  {Tirumkudulu}},\ }\href@noop {} {\bibfield  {journal} {\bibinfo  {journal}
  {Soft matter}\ }\textbf {\bibinfo {volume} {12}},\ \bibinfo {pages} {9402}
  (\bibinfo {year} {2016}{\natexlab{b}})}\BibitemShut {NoStop}%
\bibitem [{\citenamefont {Buscall}\ \emph {et~al.}(1988)\citenamefont
  {Buscall}, \citenamefont {Mills}, \citenamefont {Goodwin},\ and\
  \citenamefont {Lawson}}]{buscall1988scaling}%
  \BibitemOpen
  \bibfield  {author} {\bibinfo {author} {\bibfnamefont {R.}~\bibnamefont
  {Buscall}}, \bibinfo {author} {\bibfnamefont {P.~D.}\ \bibnamefont {Mills}},
  \bibinfo {author} {\bibfnamefont {J.~W.}\ \bibnamefont {Goodwin}}, \ and\
  \bibinfo {author} {\bibfnamefont {D.}~\bibnamefont {Lawson}},\ }\href@noop {}
  {\bibfield  {journal} {\bibinfo  {journal} {Journal of the Chemical Society,
  Faraday Transactions 1: Physical Chemistry in Condensed Phases}\ }\textbf
  {\bibinfo {volume} {84}},\ \bibinfo {pages} {4249} (\bibinfo {year}
  {1988})}\BibitemShut {NoStop}%
\bibitem [{\citenamefont {Islam}\ and\ \citenamefont
  {Lester}(2021{\natexlab{a}})}]{islam2021normal}%
  \BibitemOpen
  \bibfield  {author} {\bibinfo {author} {\bibfnamefont {M.~M.}\ \bibnamefont
  {Islam}}\ and\ \bibinfo {author} {\bibfnamefont {D.~R.}\ \bibnamefont
  {Lester}},\ }\href@noop {} {\bibfield  {journal} {\bibinfo  {journal}
  {Rheologica Acta}\ }\textbf {\bibinfo {volume} {60}},\ \bibinfo {pages} {59}
  (\bibinfo {year} {2021}{\natexlab{a}})}\BibitemShut {NoStop}%
\bibitem [{\citenamefont {Seto}\ \emph
  {et~al.}(2013{\natexlab{a}})\citenamefont {Seto}, \citenamefont {Botet},
  \citenamefont {Meireles}, \citenamefont {Auernhammer},\ and\ \citenamefont
  {Cabane}}]{seto2013compressive}%
  \BibitemOpen
  \bibfield  {author} {\bibinfo {author} {\bibfnamefont {R.}~\bibnamefont
  {Seto}}, \bibinfo {author} {\bibfnamefont {R.}~\bibnamefont {Botet}},
  \bibinfo {author} {\bibfnamefont {M.}~\bibnamefont {Meireles}}, \bibinfo
  {author} {\bibfnamefont {G.~K.}\ \bibnamefont {Auernhammer}}, \ and\ \bibinfo
  {author} {\bibfnamefont {B.}~\bibnamefont {Cabane}},\ }\href@noop {}
  {\bibfield  {journal} {\bibinfo  {journal} {Journal of rheology}\ }\textbf
  {\bibinfo {volume} {57}},\ \bibinfo {pages} {1347} (\bibinfo {year}
  {2013}{\natexlab{a}})}\BibitemShut {NoStop}%
\bibitem [{\citenamefont {Plimpton}(1995)}]{plimpton1995fast}%
  \BibitemOpen
  \bibfield  {author} {\bibinfo {author} {\bibfnamefont {S.}~\bibnamefont
  {Plimpton}},\ }\href@noop {} {\bibfield  {journal} {\bibinfo  {journal}
  {Journal of computational physics}\ }\textbf {\bibinfo {volume} {117}},\
  \bibinfo {pages} {1} (\bibinfo {year} {1995})}\BibitemShut {NoStop}%
\bibitem [{\citenamefont {Brambilla}\ \emph {et~al.}(2011)\citenamefont
  {Brambilla}, \citenamefont {Buzzaccaro}, \citenamefont {Piazza},
  \citenamefont {Berthier},\ and\ \citenamefont
  {Cipelletti}}]{brambilla2011highly}%
  \BibitemOpen
  \bibfield  {author} {\bibinfo {author} {\bibfnamefont {G.}~\bibnamefont
  {Brambilla}}, \bibinfo {author} {\bibfnamefont {S.}~\bibnamefont
  {Buzzaccaro}}, \bibinfo {author} {\bibfnamefont {R.}~\bibnamefont {Piazza}},
  \bibinfo {author} {\bibfnamefont {L.}~\bibnamefont {Berthier}}, \ and\
  \bibinfo {author} {\bibfnamefont {L.}~\bibnamefont {Cipelletti}},\
  }\href@noop {} {\bibfield  {journal} {\bibinfo  {journal} {Physical review
  letters}\ }\textbf {\bibinfo {volume} {106}},\ \bibinfo {pages} {118302}
  (\bibinfo {year} {2011})}\BibitemShut {NoStop}%
\bibitem [{\citenamefont {Lin}\ \emph {et~al.}(2015)\citenamefont {Lin},
  \citenamefont {Guy}, \citenamefont {Hermes}, \citenamefont {Ness},
  \citenamefont {Sun}, \citenamefont {Poon},\ and\ \citenamefont
  {Cohen}}]{lin2015hydrodynamic}%
  \BibitemOpen
  \bibfield  {author} {\bibinfo {author} {\bibfnamefont {N.~Y.}\ \bibnamefont
  {Lin}}, \bibinfo {author} {\bibfnamefont {B.~M.}\ \bibnamefont {Guy}},
  \bibinfo {author} {\bibfnamefont {M.}~\bibnamefont {Hermes}}, \bibinfo
  {author} {\bibfnamefont {C.}~\bibnamefont {Ness}}, \bibinfo {author}
  {\bibfnamefont {J.}~\bibnamefont {Sun}}, \bibinfo {author} {\bibfnamefont
  {W.~C.}\ \bibnamefont {Poon}}, \ and\ \bibinfo {author} {\bibfnamefont
  {I.}~\bibnamefont {Cohen}},\ }\href@noop {} {\bibfield  {journal} {\bibinfo
  {journal} {Physical review letters}\ }\textbf {\bibinfo {volume} {115}},\
  \bibinfo {pages} {228304} (\bibinfo {year} {2015})}\BibitemShut {NoStop}%
\bibitem [{\citenamefont {Colombo}\ and\ \citenamefont
  {Del~Gado}(2014)}]{colombo2014stress}%
  \BibitemOpen
  \bibfield  {author} {\bibinfo {author} {\bibfnamefont {J.}~\bibnamefont
  {Colombo}}\ and\ \bibinfo {author} {\bibfnamefont {E.}~\bibnamefont
  {Del~Gado}},\ }\href@noop {} {\bibfield  {journal} {\bibinfo  {journal}
  {Journal of rheology}\ }\textbf {\bibinfo {volume} {58}},\ \bibinfo {pages}
  {1089} (\bibinfo {year} {2014})}\BibitemShut {NoStop}%
\bibitem [{\citenamefont {Seto}\ \emph
  {et~al.}(2013{\natexlab{b}})\citenamefont {Seto}, \citenamefont {Mari},
  \citenamefont {Morris},\ and\ \citenamefont {Denn}}]{seto2013discontinuous}%
  \BibitemOpen
  \bibfield  {author} {\bibinfo {author} {\bibfnamefont {R.}~\bibnamefont
  {Seto}}, \bibinfo {author} {\bibfnamefont {R.}~\bibnamefont {Mari}}, \bibinfo
  {author} {\bibfnamefont {J.~F.}\ \bibnamefont {Morris}}, \ and\ \bibinfo
  {author} {\bibfnamefont {M.~M.}\ \bibnamefont {Denn}},\ }\href@noop {}
  {\bibfield  {journal} {\bibinfo  {journal} {Physical review letters}\
  }\textbf {\bibinfo {volume} {111}},\ \bibinfo {pages} {218301} (\bibinfo
  {year} {2013}{\natexlab{b}})}\BibitemShut {NoStop}%
\bibitem [{\citenamefont {Hsu}\ \emph {et~al.}(2018)\citenamefont {Hsu},
  \citenamefont {Ramakrishna}, \citenamefont {Zanini}, \citenamefont
  {Spencer},\ and\ \citenamefont {Isa}}]{hsu2018roughness}%
  \BibitemOpen
  \bibfield  {author} {\bibinfo {author} {\bibfnamefont {C.-P.}\ \bibnamefont
  {Hsu}}, \bibinfo {author} {\bibfnamefont {S.~N.}\ \bibnamefont
  {Ramakrishna}}, \bibinfo {author} {\bibfnamefont {M.}~\bibnamefont {Zanini}},
  \bibinfo {author} {\bibfnamefont {N.~D.}\ \bibnamefont {Spencer}}, \ and\
  \bibinfo {author} {\bibfnamefont {L.}~\bibnamefont {Isa}},\ }\href@noop {}
  {\bibfield  {journal} {\bibinfo  {journal} {Proceedings of the National
  Academy of Sciences}\ }\textbf {\bibinfo {volume} {115}},\ \bibinfo {pages}
  {5117} (\bibinfo {year} {2018})}\BibitemShut {NoStop}%
\bibitem [{\citenamefont {Mari}\ \emph {et~al.}(2015)\citenamefont {Mari},
  \citenamefont {Seto}, \citenamefont {Morris},\ and\ \citenamefont
  {Denn}}]{mari2015discontinuous}%
  \BibitemOpen
  \bibfield  {author} {\bibinfo {author} {\bibfnamefont {R.}~\bibnamefont
  {Mari}}, \bibinfo {author} {\bibfnamefont {R.}~\bibnamefont {Seto}}, \bibinfo
  {author} {\bibfnamefont {J.~F.}\ \bibnamefont {Morris}}, \ and\ \bibinfo
  {author} {\bibfnamefont {M.~M.}\ \bibnamefont {Denn}},\ }\href@noop {}
  {\bibfield  {journal} {\bibinfo  {journal} {Proceedings of the National
  Academy of Sciences}\ }\textbf {\bibinfo {volume} {112}},\ \bibinfo {pages}
  {15326} (\bibinfo {year} {2015})}\BibitemShut {NoStop}%
\bibitem [{\citenamefont {Singh}\ \emph {et~al.}(2020)\citenamefont {Singh},
  \citenamefont {Ness}, \citenamefont {Seto}, \citenamefont {de~Pablo},\ and\
  \citenamefont {Jaeger}}]{singh2020shear}%
  \BibitemOpen
  \bibfield  {author} {\bibinfo {author} {\bibfnamefont {A.}~\bibnamefont
  {Singh}}, \bibinfo {author} {\bibfnamefont {C.}~\bibnamefont {Ness}},
  \bibinfo {author} {\bibfnamefont {R.}~\bibnamefont {Seto}}, \bibinfo {author}
  {\bibfnamefont {J.~J.}\ \bibnamefont {de~Pablo}}, \ and\ \bibinfo {author}
  {\bibfnamefont {H.~M.}\ \bibnamefont {Jaeger}},\ }\href@noop {} {\bibfield
  {journal} {\bibinfo  {journal} {Physical Review Letters}\ }\textbf {\bibinfo
  {volume} {124}},\ \bibinfo {pages} {248005} (\bibinfo {year}
  {2020})}\BibitemShut {NoStop}%
\bibitem [{\citenamefont {Roy}\ and\ \citenamefont
  {Tirumkudulu}(2016{\natexlab{c}})}]{roy2016yielding}%
  \BibitemOpen
  \bibfield  {author} {\bibinfo {author} {\bibfnamefont {S.}~\bibnamefont
  {Roy}}\ and\ \bibinfo {author} {\bibfnamefont {M.~S.}\ \bibnamefont
  {Tirumkudulu}},\ }\href@noop {} {\bibfield  {journal} {\bibinfo  {journal}
  {Journal of Rheology}\ }\textbf {\bibinfo {volume} {60}},\ \bibinfo {pages}
  {559} (\bibinfo {year} {2016}{\natexlab{c}})}\BibitemShut {NoStop}%
\bibitem [{\citenamefont {Islam}\ and\ \citenamefont
  {Lester}(2021{\natexlab{b}})}]{islam2021consolidation}%
  \BibitemOpen
  \bibfield  {author} {\bibinfo {author} {\bibfnamefont {M.~M.}\ \bibnamefont
  {Islam}}\ and\ \bibinfo {author} {\bibfnamefont {D.~R.}\ \bibnamefont
  {Lester}},\ }\href@noop {} {\bibfield  {journal} {\bibinfo  {journal} {Soft
  Matter}\ }\textbf {\bibinfo {volume} {17}},\ \bibinfo {pages} {2242}
  (\bibinfo {year} {2021}{\natexlab{b}})}\BibitemShut {NoStop}%
\bibitem [{\citenamefont {Islam}\ and\ \citenamefont
  {Lester}(2021{\natexlab{c}})}]{islam2021superposed}%
  \BibitemOpen
  \bibfield  {author} {\bibinfo {author} {\bibfnamefont {M.}~\bibnamefont
  {Islam}}\ and\ \bibinfo {author} {\bibfnamefont {D.}~\bibnamefont {Lester}},\
  }\href@noop {} {\bibfield  {journal} {\bibinfo  {journal} {Journal of
  Rheology}\ }\textbf {\bibinfo {volume} {65}},\ \bibinfo {pages} {837}
  (\bibinfo {year} {2021}{\natexlab{c}})}\BibitemShut {NoStop}%
\bibitem [{\citenamefont {Luding}(2008)}]{luding2008cohesive}%
  \BibitemOpen
  \bibfield  {author} {\bibinfo {author} {\bibfnamefont {S.}~\bibnamefont
  {Luding}},\ }\href@noop {} {\bibfield  {journal} {\bibinfo  {journal}
  {Granular matter}\ }\textbf {\bibinfo {volume} {10}},\ \bibinfo {pages} {235}
  (\bibinfo {year} {2008})}\BibitemShut {NoStop}%
\bibitem [{\citenamefont {Landau}\ \emph {et~al.}(1986)\citenamefont {Landau},
  \citenamefont {Lif{\v{s}}ic}, \citenamefont {Lifshitz}, \citenamefont
  {Kosevich},\ and\ \citenamefont {Pitaevskii}}]{landau1986theory}%
  \BibitemOpen
  \bibfield  {author} {\bibinfo {author} {\bibfnamefont {L.~D.}\ \bibnamefont
  {Landau}}, \bibinfo {author} {\bibfnamefont {E.~M.}\ \bibnamefont
  {Lif{\v{s}}ic}}, \bibinfo {author} {\bibfnamefont {E.~M.}\ \bibnamefont
  {Lifshitz}}, \bibinfo {author} {\bibfnamefont {A.~M.}\ \bibnamefont
  {Kosevich}}, \ and\ \bibinfo {author} {\bibfnamefont {L.~P.}\ \bibnamefont
  {Pitaevskii}},\ }\href@noop {} {\emph {\bibinfo {title} {Theory of
  elasticity: volume 7}}},\ Vol.~\bibinfo {volume} {7}\ (\bibinfo  {publisher}
  {Elsevier},\ \bibinfo {year} {1986})\BibitemShut {NoStop}%
\bibitem [{\citenamefont {Rothenburg}\ and\ \citenamefont
  {Bathurst}(1989)}]{rothenburg1989analytical}%
  \BibitemOpen
  \bibfield  {author} {\bibinfo {author} {\bibfnamefont {L.}~\bibnamefont
  {Rothenburg}}\ and\ \bibinfo {author} {\bibfnamefont {R.}~\bibnamefont
  {Bathurst}},\ }\href@noop {} {\bibfield  {journal} {\bibinfo  {journal}
  {Geotechnique}\ }\textbf {\bibinfo {volume} {39}},\ \bibinfo {pages} {601}
  (\bibinfo {year} {1989})}\BibitemShut {NoStop}%
\bibitem [{\citenamefont {Liu}\ \emph {et~al.}(1995)\citenamefont {Liu},
  \citenamefont {Nagel}, \citenamefont {Schecter}, \citenamefont {Coppersmith},
  \citenamefont {Majumdar}, \citenamefont {Narayan},\ and\ \citenamefont
  {Witten}}]{liu1995force}%
  \BibitemOpen
  \bibfield  {author} {\bibinfo {author} {\bibfnamefont {C.-h.}\ \bibnamefont
  {Liu}}, \bibinfo {author} {\bibfnamefont {S.~R.}\ \bibnamefont {Nagel}},
  \bibinfo {author} {\bibfnamefont {D.}~\bibnamefont {Schecter}}, \bibinfo
  {author} {\bibfnamefont {S.}~\bibnamefont {Coppersmith}}, \bibinfo {author}
  {\bibfnamefont {S.}~\bibnamefont {Majumdar}}, \bibinfo {author}
  {\bibfnamefont {O.}~\bibnamefont {Narayan}}, \ and\ \bibinfo {author}
  {\bibfnamefont {T.}~\bibnamefont {Witten}},\ }\href@noop {} {\bibfield
  {journal} {\bibinfo  {journal} {Science}\ }\textbf {\bibinfo {volume}
  {269}},\ \bibinfo {pages} {513} (\bibinfo {year} {1995})}\BibitemShut
  {NoStop}%
\bibitem [{\citenamefont {Mueth}\ \emph {et~al.}(1998)\citenamefont {Mueth},
  \citenamefont {Jaeger},\ and\ \citenamefont {Nagel}}]{mueth1998force}%
  \BibitemOpen
  \bibfield  {author} {\bibinfo {author} {\bibfnamefont {D.~M.}\ \bibnamefont
  {Mueth}}, \bibinfo {author} {\bibfnamefont {H.~M.}\ \bibnamefont {Jaeger}}, \
  and\ \bibinfo {author} {\bibfnamefont {S.~R.}\ \bibnamefont {Nagel}},\
  }\href@noop {} {\bibfield  {journal} {\bibinfo  {journal} {Physical Review
  E}\ }\textbf {\bibinfo {volume} {57}},\ \bibinfo {pages} {3164} (\bibinfo
  {year} {1998})}\BibitemShut {NoStop}%
\bibitem [{\citenamefont {Majmudar}\ and\ \citenamefont
  {Behringer}(2005)}]{majmudar2005contact}%
  \BibitemOpen
  \bibfield  {author} {\bibinfo {author} {\bibfnamefont {T.~S.}\ \bibnamefont
  {Majmudar}}\ and\ \bibinfo {author} {\bibfnamefont {R.~P.}\ \bibnamefont
  {Behringer}},\ }\href@noop {} {\bibfield  {journal} {\bibinfo  {journal}
  {nature}\ }\textbf {\bibinfo {volume} {435}},\ \bibinfo {pages} {1079}
  (\bibinfo {year} {2005})}\BibitemShut {NoStop}%
\bibitem [{\citenamefont {Radjai}\ \emph {et~al.}(1996)\citenamefont {Radjai},
  \citenamefont {Jean}, \citenamefont {Moreau},\ and\ \citenamefont
  {Roux}}]{radjai1996force}%
  \BibitemOpen
  \bibfield  {author} {\bibinfo {author} {\bibfnamefont {F.}~\bibnamefont
  {Radjai}}, \bibinfo {author} {\bibfnamefont {M.}~\bibnamefont {Jean}},
  \bibinfo {author} {\bibfnamefont {J.-J.}\ \bibnamefont {Moreau}}, \ and\
  \bibinfo {author} {\bibfnamefont {S.}~\bibnamefont {Roux}},\ }\href@noop {}
  {\bibfield  {journal} {\bibinfo  {journal} {Physical review letters}\
  }\textbf {\bibinfo {volume} {77}},\ \bibinfo {pages} {274} (\bibinfo {year}
  {1996})}\BibitemShut {NoStop}%
\bibitem [{\citenamefont {O'Hern}\ \emph {et~al.}(2002)\citenamefont {O'Hern},
  \citenamefont {Langer}, \citenamefont {Liu},\ and\ \citenamefont
  {Nagel}}]{o2002random}%
  \BibitemOpen
  \bibfield  {author} {\bibinfo {author} {\bibfnamefont {C.~S.}\ \bibnamefont
  {O'Hern}}, \bibinfo {author} {\bibfnamefont {S.~A.}\ \bibnamefont {Langer}},
  \bibinfo {author} {\bibfnamefont {A.~J.}\ \bibnamefont {Liu}}, \ and\
  \bibinfo {author} {\bibfnamefont {S.~R.}\ \bibnamefont {Nagel}},\ }\href@noop
  {} {\bibfield  {journal} {\bibinfo  {journal} {Physical Review Letters}\
  }\textbf {\bibinfo {volume} {88}},\ \bibinfo {pages} {075507} (\bibinfo
  {year} {2002})}\BibitemShut {NoStop}%
\bibitem [{\citenamefont {Akella}\ \emph {et~al.}(2018)\citenamefont {Akella},
  \citenamefont {Bandi}, \citenamefont {Hentschel}, \citenamefont {Procaccia},\
  and\ \citenamefont {Roy}}]{akella2018force}%
  \BibitemOpen
  \bibfield  {author} {\bibinfo {author} {\bibfnamefont {V.}~\bibnamefont
  {Akella}}, \bibinfo {author} {\bibfnamefont {M.}~\bibnamefont {Bandi}},
  \bibinfo {author} {\bibfnamefont {H.~G.~E.}\ \bibnamefont {Hentschel}},
  \bibinfo {author} {\bibfnamefont {I.}~\bibnamefont {Procaccia}}, \ and\
  \bibinfo {author} {\bibfnamefont {S.}~\bibnamefont {Roy}},\ }\href@noop {}
  {\bibfield  {journal} {\bibinfo  {journal} {Physical Review E}\ }\textbf
  {\bibinfo {volume} {98}},\ \bibinfo {pages} {012905} (\bibinfo {year}
  {2018})}\BibitemShut {NoStop}%
\bibitem [{\citenamefont {Radjai}\ \emph {et~al.}(1998)\citenamefont {Radjai},
  \citenamefont {Wolf}, \citenamefont {Jean},\ and\ \citenamefont
  {Moreau}}]{radjai1998bimodal}%
  \BibitemOpen
  \bibfield  {author} {\bibinfo {author} {\bibfnamefont {F.}~\bibnamefont
  {Radjai}}, \bibinfo {author} {\bibfnamefont {D.~E.}\ \bibnamefont {Wolf}},
  \bibinfo {author} {\bibfnamefont {M.}~\bibnamefont {Jean}}, \ and\ \bibinfo
  {author} {\bibfnamefont {J.-J.}\ \bibnamefont {Moreau}},\ }\href@noop {}
  {\bibfield  {journal} {\bibinfo  {journal} {Physical review letters}\
  }\textbf {\bibinfo {volume} {80}},\ \bibinfo {pages} {61} (\bibinfo {year}
  {1998})}\BibitemShut {NoStop}%
\bibitem [{\citenamefont {Richefeu}\ \emph {et~al.}(2009)\citenamefont
  {Richefeu}, \citenamefont {El~Youssoufi}, \citenamefont {Az{\'e}ma},\ and\
  \citenamefont {Radjai}}]{richefeu2009force}%
  \BibitemOpen
  \bibfield  {author} {\bibinfo {author} {\bibfnamefont {V.}~\bibnamefont
  {Richefeu}}, \bibinfo {author} {\bibfnamefont {M.~S.}\ \bibnamefont
  {El~Youssoufi}}, \bibinfo {author} {\bibfnamefont {E.}~\bibnamefont
  {Az{\'e}ma}}, \ and\ \bibinfo {author} {\bibfnamefont {F.}~\bibnamefont
  {Radjai}},\ }\href@noop {} {\bibfield  {journal} {\bibinfo  {journal} {Powder
  Technology}\ }\textbf {\bibinfo {volume} {190}},\ \bibinfo {pages} {258}
  (\bibinfo {year} {2009})}\BibitemShut {NoStop}%
\bibitem [{\citenamefont {Richefeu}\ \emph
  {et~al.}(2006{\natexlab{a}})\citenamefont {Richefeu}, \citenamefont
  {Radja{\i}},\ and\ \citenamefont {El~Youssoufi}}]{richefeu2006stress}%
  \BibitemOpen
  \bibfield  {author} {\bibinfo {author} {\bibfnamefont {V.}~\bibnamefont
  {Richefeu}}, \bibinfo {author} {\bibfnamefont {F.}~\bibnamefont {Radja{\i}}},
  \ and\ \bibinfo {author} {\bibfnamefont {M.~S.}\ \bibnamefont
  {El~Youssoufi}},\ }\href@noop {} {\bibfield  {journal} {\bibinfo  {journal}
  {The European Physical Journal E}\ }\textbf {\bibinfo {volume} {21}},\
  \bibinfo {pages} {359} (\bibinfo {year} {2006}{\natexlab{a}})}\BibitemShut
  {NoStop}%
\bibitem [{\citenamefont {Richefeu}\ \emph
  {et~al.}(2006{\natexlab{b}})\citenamefont {Richefeu}, \citenamefont
  {El~Youssoufi},\ and\ \citenamefont {Radjai}}]{richefeu2006shear}%
  \BibitemOpen
  \bibfield  {author} {\bibinfo {author} {\bibfnamefont {V.}~\bibnamefont
  {Richefeu}}, \bibinfo {author} {\bibfnamefont {M.~S.}\ \bibnamefont
  {El~Youssoufi}}, \ and\ \bibinfo {author} {\bibfnamefont {F.}~\bibnamefont
  {Radjai}},\ }\href@noop {} {\bibfield  {journal} {\bibinfo  {journal}
  {Physical Review E}\ }\textbf {\bibinfo {volume} {73}},\ \bibinfo {pages}
  {051304} (\bibinfo {year} {2006}{\natexlab{b}})}\BibitemShut {NoStop}%
\bibitem [{\citenamefont {Gilabert}\ \emph {et~al.}(2007)\citenamefont
  {Gilabert}, \citenamefont {Roux},\ and\ \citenamefont
  {Castellanos}}]{gilabert2007computer}%
  \BibitemOpen
  \bibfield  {author} {\bibinfo {author} {\bibfnamefont {F.}~\bibnamefont
  {Gilabert}}, \bibinfo {author} {\bibfnamefont {J.-N.}\ \bibnamefont {Roux}},
  \ and\ \bibinfo {author} {\bibfnamefont {A.}~\bibnamefont {Castellanos}},\
  }\href@noop {} {\bibfield  {journal} {\bibinfo  {journal} {Physical review
  E}\ }\textbf {\bibinfo {volume} {75}},\ \bibinfo {pages} {011303} (\bibinfo
  {year} {2007})}\BibitemShut {NoStop}%
\bibitem [{\citenamefont {Dinsmore}\ and\ \citenamefont
  {Weitz}(2002)}]{dinsmore2002direct}%
  \BibitemOpen
  \bibfield  {author} {\bibinfo {author} {\bibfnamefont {A.}~\bibnamefont
  {Dinsmore}}\ and\ \bibinfo {author} {\bibfnamefont {D.}~\bibnamefont
  {Weitz}},\ }\href@noop {} {\bibfield  {journal} {\bibinfo  {journal} {Journal
  of Physics: Condensed Matter}\ }\textbf {\bibinfo {volume} {14}},\ \bibinfo
  {pages} {7581} (\bibinfo {year} {2002})}\BibitemShut {NoStop}%
\end{thebibliography}%

\end{document}